\documentclass[
 amsmath,
 amssymb,
 aps,
 prd,
 preprintnumbers,
 superscriptaddress,
 eqsecnum,
 showpacs,
 nofootinbib,
 showkeys
]{revtex4-2}
\usepackage{blkarray,graphicx,dcolumn,bm,hyperref,physics,here,mathrsfs,cases,empheq,latexsym,amsfonts,color,listings,subcaption,caption,orcidlink,comment}

\DeclareMathOperator*{\SumInt}{%
\mathchoice%
  {\ooalign{$\displaystyle\sum$\cr\hidewidth$\displaystyle\int$\hidewidth\cr}}
  {\ooalign{\raisebox{.14\height}{\scalebox{.7}{$\textstyle\sum$}}\cr\hidewidth$\textstyle\int$\hidewidth\cr}}
  {\ooalign{\raisebox{.2\height}{\scalebox{.6}{$\scriptstyle\sum$}}\cr$\scriptstyle\int$\cr}}
  {\ooalign{\raisebox{.2\height}{\scalebox{.6}{$\scriptstyle\sum$}}\cr$\scriptstyle\int$\cr}}
}

\begin{document}

\allowdisplaybreaks[1]
\preprint{YITP-24-89, IPMU24-0031}
\title{Spin wave optics for gravitational waves lensed by a Kerr black hole}

\author{Kei-ichiro Kubota\orcidlink{0000-0002-1576-4332}}
    \email{keiichiro.kubota@yukawa.kyoto-u.ac.jp}
    \affiliation{Center for Gravitational Physics, Yukawa Institute for Theoretical Physics, Kyoto University, 606-8502, Kyoto, Japan.}
\author{Shun Arai\orcidlink{0000-0002-2527-3705}}
    \affiliation{Kobayashi-Maskawa Institute for the Origin of Particles and the Universe, Nagoya University, Nagoya 464-8602, Japan}
\author{Hayato Motohashi\orcidlink{0000-0002-4330-7024}}
    \affiliation{Division of Liberal Arts, Kogakuin University, 2665-1 Nakano-machi, Hachioji, Tokyo 192-0015, Japan}
\author{Shinji Mukohyama\orcidlink{0000-0002-9934-2785}}
    \affiliation{Center for Gravitational Physics, Yukawa Institute for Theoretical Physics, Kyoto University, 606-8502, Kyoto, Japan.}
    \affiliation{Kavli Institute for the Physics and Mathematics of the Universe (WPI), The University of Tokyo Institutes for Advanced Study, The University of Tokyo, Kashiwa, Chiba, 277-8583, Japan.}
\date{\today}
\begin{abstract}
    Gravitational waves exhibit the unique signature of their spin-2 nature in processes of wave scattering, due to the interaction between spin and a background spacetime.
    Since the spin effect is more pronounced for longer wavelengths 
    and gravitational waves sourced by binaries have a long wavelength, it may become an important effect in addition to the wave effect.
    We study the propagation of gravitational waves lensed by a Kerr black hole by numerically solving the Teukolsky equation with a source term of the equal-mass circular binary, taking into account both spin effect and wave effect. We find helicity-dependent small-period oscillation in the power spectrum of the amplification factor in the forward direction and the oscillation is enhanced as spin of a prograde Kerr black hole increases.
\end{abstract}
\maketitle

\section{\label{sec:intro}Introduction}
Gravitational lensing of gravitational waves has been investigated as a probe for cosmology and astrophysics.
One particularly intriguing aspect of this phenomenon is the wave effect (see Ref.~\cite{Leung:2023lmq} for a review).
For instance, gravitational waves sourced by a massive black hole binary have long wavelengths comparable to the Schwarzschild radius of supermassive black holes, making the wave effect significant.
Recently, the wave effect for gravitational waves has garnered significant attention because it enables the estimation of the lens parameter~\cite{Takahashi:2003ix,Caliskan:2022hbu,Tambalo:2022wlm} and the small scale matter power spectrum~\cite{Jung:2017flg,Diego:2019lcd,Yeung:2021chy,Guo:2022dre,Fairbairn:2022xln,Savastano:2023spl,Urrutia:2024pos}.
In other cases, the wave effect in gravitational lensing has also been investigated for a point mass~\cite{Deguchi:1986zz,Schneider:1992bmb,Nakamura:1999uwi}~\footnote{It is worth mentioning that the gravitational lensing by a Schwarzschild black hole, without the weak gravitational field assumption, has been analyzed within the framework of geometric optics~\cite{Virbhadra:1999nm}.}, a rotating object~\cite{Baraldo:1999ny}, a binary system~\cite{Mehrabi:2012dy}, a cosmic string~\cite{Suyama:2005ez,Yoo:2012dn}, and an Ellis wormhole~\cite{Yoo:2013cia}.
Additionally, the wave effect has also been investigated in the context of black hole shadow~\cite{Nambu:2015aea,Nambu:2019sqn,Willenborg:2023ixu}.
An interesting result of Ref.~\cite{Nambu:2019sqn} is the appearance of small-period oscillation in the power spectrum of the scalar field in the forward direction due to interference between the direct ray and the winding ray. 
An improved calculation also confirms this oscillation~\cite{Motohashi:2021zyv}.
The oscillation is related to the photon sphere and thus captures information about the strong gravitational field.

In addition to the wave property, a characteristic property of gravitational waves is that they are spin-2 fields.
It has long been known that the interaction between the spin of the particle and the background spacetime affects the motion of the particle.
Various approaches to the motion of massless spinning particles such as the Souriau-Saturunini equations~\cite{AIHPA_1974__20_4_315_0,saturnini:tel-01344863,Duval:2018hzh,Duval:2016hxo}, the quantum mechanical approach~\cite{Berard:2004xn,Gosselin:2006wp,Yamamoto:2017gla,Oancea:2020khc,Oancea:2022szu,Andersson:2023bvw,Andersson:2020gsj}, and the spin-optics approach~\cite{Frolov:2011mh,Yoo:2012vv,Frolov:2012zn,Shoom:2020zhr,Frolov:2020uhn,Dahal:2022gop,Dahal:2022gop,Harte:2022dpo,Kubota:2023dlz,Oancea:2023hgu,Frolov:2024ebe,Frolov:2024} (see Ref.~\cite{Oancea:2019pgm} for the relation between them) have been developed.
In these approaches, the spin effect appears as a correction to the short-wavelength limit, becoming more pronounced for longer wavelengths.
Therefore, in addition to the wave effect, the spin effect may also be a key effect in the propagation of long-wavelength gravitational waves.
By taking the spin effect into account, we could obtain additional information that cannot be captured by scalar waves, such as the rotation of lens black hole.
However, most studies on wave optics except for recent papers~\cite{Braga:2024pik,Pijnenburg:2024btj} neglect the polarization tensor, treating gravitational waves as scalar waves.

The approach of treating gravitational waves as scalar waves can only be applied to gravitational waves with wavelengths shorter than the radius of curvature of the background spacetime.
For longer wavelengths, it might be necessary to account for the spin effect in addition to the wave effect.
In order to investigate the propagation taking both wave effect and spin effect into account in the Kerr spacetime, Teukolsky formalism can be employed.
The scattering of plane gravitational waves with long wavelengths has been extensively studied using the Teukolsky equation~\cite{Chrzanowski:1976jb,DeLogi:1977dp,Handler:1980un,Futtermanthesis,Dolan:2008kf}.
When we consider the lensing of gravitational waves sourced by a binary system, backward gravitational waves are calculated with the differential scattering cross section, which is the result of the studies of the plane wave scattering.
This is because only the path with the small impact parameter contributes, making the plane wave approximation applicable.
However, the forward gravitational waves could not be calculated with only the differential scattering cross section because contributions arise from both small and large impact parameter paths, making the plane wave approximation inapplicable.
Indeed, the differential scattering cross section in the exact forward direction diverges due to the long-range nature of the gravitational field, i.e., $\mathrm{d}\sigma(0)/\mathrm{d}\Omega = \infty $~\cite{Dolan:2008kf}, and thus the calculation only with the differential scattering cross section cannot capture the physics of the forward gravitational waves.
The spin effect on the forward long-wavelength gravitational waves has been poorly studied.

In this paper, we explore spin wave optics of gravitational waves.
We investigate the spin effect on the lensed gravitational waves by numerically solving the Teukolsky equation with a binary system as a source term, without relying on the differential scattering cross section.

The rest of this paper is organized as follows. In Sec.~\ref{sec:teukolsky},
we give a brief review of the Teukolsky formalism.
We explain the method of the numerical calculation of gravitational waves emitted from an equal-mass circular binary and then lensed by a Kerr black hole and show the numerical result in Sec.~\ref{sec:lensing}. 
Finally, we conclude the paper in Sec.~\ref{sec:conclusion}.
We use the unit $c=G=1$.

\section{\label{sec:teukolsky}Teukolsky Formalism for gravitational waves}
We briefly review the Teukolsky formalism based on Ref.~\cite{Sasaki:2003xr}, describing the basic equations for wave scattering of gravitational waves.
The Kerr metric in the Boyer-Lindquist coordinate is given by
\begin{align}
    \mathrm{d}s^2 =& -\frac{\Delta}{\Sigma}(\mathrm{d}t-a\sin^2\theta\mathrm{d}\varphi)^2+\frac{\sin^2\theta}{\Sigma}\left((r^2+a^2)\mathrm{d}\varphi-a\mathrm{d}t\right)^2 +\frac{\Sigma}{\Delta}\mathrm{d}r^2 + \Sigma \mathrm{d}\theta^2,
    \label{eq:metric}
\end{align}
where 
\begin{equation}
    \Sigma \coloneqq r^2 + a^2 \cos^2\theta,\quad 
    \Delta \coloneqq r^2 - 2Mr + a^2 = (r - r_+)(r - r_-),\quad 
    r_{\pm} \coloneqq M \pm \sqrt{M^2 - a^2}, 
\end{equation}
and $M$ and $a$ denote the mass and the spin parameter of a Kerr black hole, respectively. 
The Teukolsky equation with a source term is given by
\begin{align}
    \hat{\mathcal{O}}_s\psi_s =& 4\pi \Sigma \hat{T}_s,
\end{align}
where $s$ denotes the spin weight of a wave, and $\psi_s$ and $\hat{T}_s$ are the spin-weighted variable and the source term for the spin $s$, respectively. The second-order differential operator $\hat{\mathcal{O}}_s$ is defined as
\begin{align}
    \hat{\mathcal{O}}_s \coloneqq& \Biggl[ -\left(\frac{(r^2+a^2)^2}{\Delta}-a^2\sin^2\theta\right)\partial_t^2-\frac{4Mar}{\Delta}\partial_t\partial_\varphi+\left(\frac{1}{\sin^2\theta}-\frac{a^2}{\Delta}\right)\partial_\varphi^2+\Delta^{-s}\partial_r(\Delta^{s+1}\partial_r)+\frac{1}{\sin\theta}\partial_\theta(\sin\theta\partial_\theta)\nonumber \\
    &+2s\left(\frac{a(r-M)}{\Delta}+\frac{i\cos\theta}{\sin^2\theta}\right)\partial_\varphi+2s\left(\frac{M(r^2-a^2)}{\Delta}-r-ia\cos\theta\right)\partial_t+s-s^2\cot^2\theta\Biggr].
\end{align}
Throughout the paper, we focus on the scattering of gravitational waves, setting $s = -2$. We represent $\psi_{-2} = \rho^{-4} \Psi_4$ with $\rho \coloneqq (r - i a \cos\theta)^{-1}$ in terms of the complex scalar in Newman-Penrose formalism: $\Psi_4 = -C_{\mu\nu\rho\sigma} n^\mu \bar{m}^\nu n^\rho \bar{m}^\sigma$, where $C_{\mu\nu\rho\sigma}$ is the Weyl tensor and
\begin{align}
    n^\alpha \coloneqq \frac{1}{2\Sigma}\begin{pmatrix} r^2+a^2, & -\Delta, & 0, & a\end{pmatrix},\quad
    m^\alpha \coloneqq \frac{1}{\sqrt{2}(r+ia\cos\theta)}\begin{pmatrix} ia\sin\theta, & 0, & 1, & \displaystyle\frac{i}{\sin\theta}\end{pmatrix},
\end{align}
together with another independent null vector, define the null tetrad. 

\begin{figure}[h]
    \includegraphics[width=0.3\columnwidth]{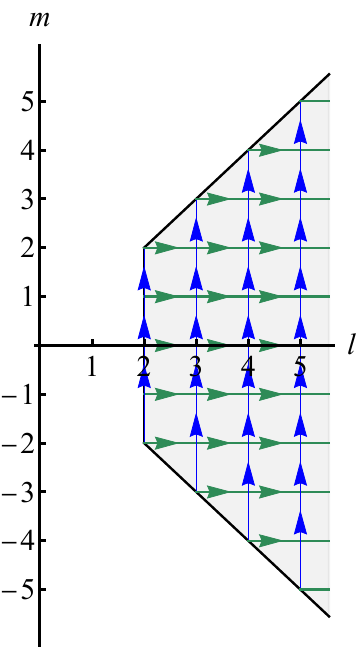}
    \caption{\label{fig:lsummsum}Exchange of the summation over $l$ and the summation over $m$.}
\end{figure}

The source term for $s=-2$~\footnote{Note that a prime is not a derivative operator but a GHP operator~\cite{Geroch:1973am}.}~\footnote{There is a typo in Eq. (8) in the Living Reviews article~\cite{Sasaki:2003xr}. $\rho^{-4}$ is deficient.} is 
\begin{align}
    \hat{T}_{-2} &\coloneqq 2\rho^{-4}(B'_2+B'^*_2),\nonumber \\
    B'_2 &\coloneqq -\frac{1}{2}\rho^8\bar{\rho}\hat{L}_{-1}\Bigl[\rho^{-4}\hat{L}_0(\rho^{-2}\bar{\rho}^{-1}T_{nn})\Bigr] - \frac{1}{2\sqrt{2}}\rho^8\bar{\rho}\Delta^2\hat{L}_{-1}\Bigl[\rho^{-4}\bar{\rho}^2\hat{J}_+(\rho^{-2}\bar{\rho}^{-2}\Delta^{-1}T_{\bar{m}n})\Bigr],\nonumber \\
    B'^*_2 &\coloneqq -\frac{1}{4}\rho^8\bar{\rho}\Delta^2\hat{J}_+\Bigl[\rho^{-4}\hat{J}_+(\rho^{-2}\bar{\rho}T_{\bar{m}\bar{m}})\Bigr] - \frac{1}{2\sqrt{2}}\rho^8\bar{\rho}\Delta^2\hat{J}_{+}\Bigl[\rho^{-4}\bar{\rho}^2\Delta^{-1}\hat{L}_{-1}(\rho^{-2}\bar{\rho}^{-2}\Delta^{-1}T_{\bar{m}n})\Bigr].
\end{align}
Here $T_{nn} = T_{\mu\nu}n^\mu n^\nu$, $T_{\bar{m}n} = T_{\mu\nu}\bar{m}^\mu n^\nu$, and $T_{\bar{m}\bar{m}} = T_{\mu\nu}\bar{m}^\mu \bar{m}^\nu$. The first-order differential operators $\hat{L}_s$ and $\hat{J}_+$ are defined as 
\begin{align}
    \hat{L}_s \coloneqq \partial_\theta -\frac{i}{\sin\theta}\partial_\varphi-ia\sin\theta\partial_t+s\cot\theta,\quad 
    \hat{J}_+ \coloneqq \partial_r -\frac{1}{\Delta}((r^2+a^2)\partial_t+a\partial_\varphi).
\end{align}
Note that the Hermitian conjugate of $\hat{L}_s$ is described as $\hat{L}_s^\dagger$. The Teukolsky equation can be split into radial and angular parts.
Setting
\begin{align}
    \psi_s &= \frac{1}{\sqrt{2\pi}}\SumInt e^{-i\omega t+im\varphi}{}_{s}R_{lm\omega}(r){}_{s}S_{lm\omega}(\theta),\label{eq:psis}\\
    4\pi\Sigma\hat{T}_{s} &= \frac{1}{\sqrt{2\pi}}\SumInt e^{-i\omega t+im\varphi}{}_{s}T_{lm\omega}(r){}_{s}S_{lm\omega}(\theta),
\end{align}
the Teukolsky equation is separated into the following radial and angular parts
\begin{align}
    \Biggl[\Delta^{-s}\frac{\mathrm{d}}{\mathrm{d}r}\left(\Delta^{s+1}\frac{\mathrm{d}}{\mathrm{d}r}\right)+\frac{K^2-2is(r-M)K}{\Delta}+4is\omega r-\lambda\Biggr]{}_{s}R_{lm\omega}&={}_{s}T_{lm\omega},\label{eq:TRE}\\
    \Biggl[\frac{1}{\sin\theta}\frac{\mathrm{d}}{\mathrm{d}\theta}\left(\sin\theta\frac{\mathrm{d}}{\mathrm{d}\theta}\right)-a^2\omega^2\sin^2\theta-\frac{(m+s\cos\theta)^2}{\sin^2\theta}-2a\omega s\cos\theta+s+2a\omega m+\lambda\Biggr]{}_{s}S_{lm\omega}&=0.\label{eq:TAE}
\end{align}
Here, ${}_{s}S_{lm\omega}$ is the spin-weighted spheroidal harmonics function, $\lambda$ is the separation constant, and $K \coloneqq (r^2+a^2)\omega - ma$. The rule of summation is defined as
\begin{align}
    \SumInt &\coloneqq \sum_{l=|s|}^{\infty}\sum_{m=-l}^{m=l}\int_{-\infty}^\infty \mathrm{d}\omega = \sum_{m=-\infty}^{m=\infty}\sum_{l=\max\{|m|,|s|\}}^{\infty}\int_{-\infty}^\infty \mathrm{d}\omega. \label{eq:summation_rule}
\end{align}
Note that the second equality in Eq.~\eqref{eq:summation_rule} is derived when exchanging the summation over $\ell$ and $m$ (see Fig.~\ref{fig:lsummsum}).
Imposing the regularity of the angular function at $\theta=0$ and $\pi$, which are regular singular points, we can write ${}_{s}S_{lm\omega}(\theta)$ as
\begin{align}
    {}_{s}S_{lm\omega}(\theta)=(\theta)^{|m+s|}(\pi-\theta)^{|m-s|}\sum_{k=0}^\infty a_k \theta^k,
    \label{eq:spinweightedspheroidalharmonicsfunction}
\end{align}
where $a_k$ is a coefficient satisfying a recurrence relation obtained by plugging Eq.~\eqref{eq:spinweightedspheroidalharmonicsfunction} into the angular equation \eqref{eq:TAE}.
The regularity condition discretizes the separation constant $\lambda$.
The normalization and the orthogonality conditions of ${}_{s}S_{lm\omega}$ are respectively defined as
\begin{align}
    \int_{0}^{\pi}|{}_{s}S_{lm\omega}|^2\sin\theta\mathrm{d}\theta=1,\quad 
    \int_0^\pi \sin\theta\mathrm{d}\theta\ {}_{s}S_{lm\omega}\ {}_{s}S_{l'm\omega}=\delta_{ll'}.
\end{align}
Note that the Sturm-Liouville theory ensures that ${}_{s}S_{lm\omega}$ satisfies the orthogonal conditions for the real frequency.
Using the orthogonal conditions, the source term for the radial function is written as
\begin{align}
    {}_{-2}T_{lm\omega}
    &=\int \mathrm{d}t\frac{e^{i\omega t}}{2\pi}\int_0^{2\pi}\mathrm{d}\varphi\frac{e^{-im\varphi}}{\sqrt{2\pi}}\int_0^{\pi}\sin\theta\mathrm{d}\theta \ 4\pi\Sigma \hat{T}_{-2}\ {}_{-2}S_{lm\omega}\nonumber \\
    &=4\int \mathrm{d}t \int_0^{2\pi}\mathrm{d}\varphi \int_0^{\pi}\sin\theta\mathrm{d}\theta e^{i\omega t -im\varphi}\rho^{-5}\bar{\rho}^{-1}(B'_2+B'^{*}_2) \frac{{}_{-2}S_{lm\omega}}{\sqrt{2\pi}}.
    \label{eq:sTlmomega}
\end{align}

\begin{figure}[h]
    \includegraphics[width=0.5\columnwidth]{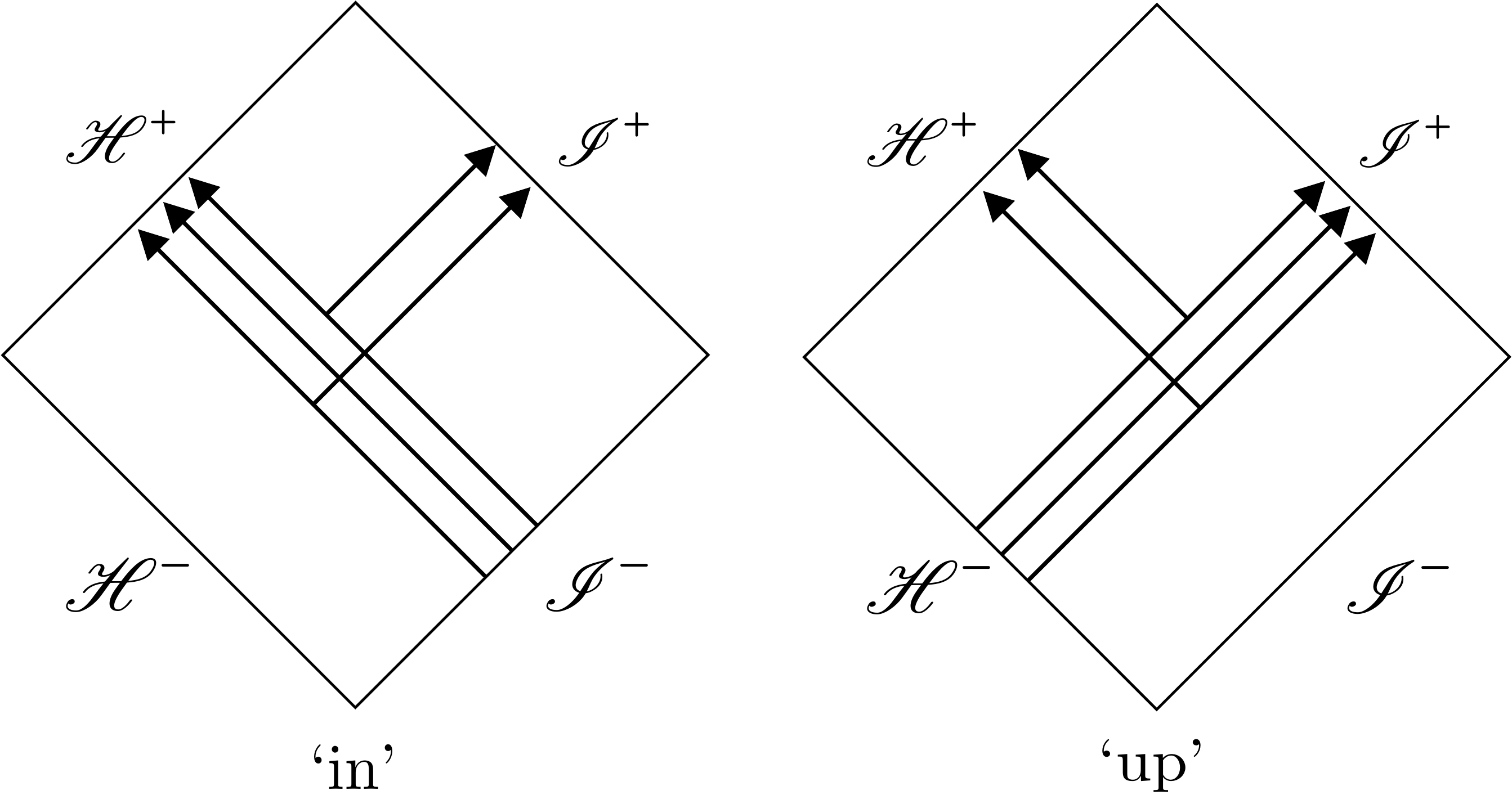}
    \caption{\label{fig:inupsolution}Conceptual illustration of the `in' and `up' solution.}
\end{figure}

The Green function of the radial Teukolsky equation~\eqref{eq:TRE} is constructed from two independent homogeneous solutions.
To this end, we introduce two independent homogeneous solutions, the `in' solution ${}_{s}R^\text{in}_{lm\omega}$ satisfying boundary conditions
\begin{align}
    R^\text{in}_{lm\omega}(r_*)\to
    \left\{ \,
        \begin{aligned}
            &B^\text{trans}_{lm\omega}\Delta^2e^{-ikr_*} & \text{for } r\to r_+,\\
            &r^3B^\text{ref}_{lm\omega}e^{i\omega r_*} + r^{-1}B^\text{inc}_{lm\omega}e^{-i\omega r_*} & \text{for } r\to \infty,
        \end{aligned}
    \right.\label{eq:Rinasymp}
\end{align}
and the `up' solution ${}_{s}R^\text{up}_{lm\omega}$ satisfying boundary conditions
\begin{align}
    R^\text{up}_{lm\omega}(r_*)\to
    \left\{ \,
        \begin{aligned}
            &C^\text{up}_{lm\omega}e^{ikr_*} + \Delta^2C^\text{ref}_{lm\omega}e^{-ik r_*}  & \text{for } r\to r_+,\\
            &r^3C^\text{trans}_{lm\omega}e^{i\omega r_*} & \text{for } r\to \infty,
        \end{aligned}
    \right.\label{eq:Rupasymp}
\end{align}
where $B^\text{trans}_{lm\omega}$, $B^\text{ref}_{lm\omega}$, $B^\text{inc}_{lm\omega}$, $C^\text{trans}_{lm\omega}$, $C^\text{ref}_{lm\omega}$, and $C^\text{up}_{lm\omega}$ are complex constants, $k = \omega - am/(2Mr_+)$, and $r_*$ is the tortoise coordinate defined by
\begin{align}
    r_*\coloneqq\int \frac{r^2+a^2}{\Delta}\mathrm{d}r=r + \frac{M}{\sqrt{M^2-a^2}}\left[r_+\ln\left(\frac{r-r_+}{2M}\right) - r_-\ln\left(\frac{r-r_-}{2M}\right)\right].
\end{align}
The `in' solution is characterized by the absence of outcoming waves from the past horizon $\mathscr{H}^{-}$, whereas the `up' solution is characterized by the absence of incoming waves at the past infinity $\mathscr{I}^{-}$ (see Fig.~\ref{fig:inupsolution}).
The radial Green function for the operator $[\cdots]$ on the left-hand side of Eq.~\eqref{eq:TRE} can be written in term of the two independent homogeneous solutions ${}_{s}R^\text{in}_{lm\omega}$ and ${}_{s}R^\text{up}_{lm\omega}$ as
\begin{align}
    {}_sG_{lm\omega}(r,r_\text{s})&= \frac{-\Delta^s(r_\text{s})}{{}_{s}\mathcal{W}_{lm\omega}}\Bigl({}_{s}R^\text{in}_{lm\omega}(r_<) {}_{s}R^\text{up}_{lm\omega}(r_>)\Bigr),\label{eq:radialGreen}
\end{align}
where $r_>=\max(r,r_\text{s})$ and $r_<=\min(r,r_\text{s})$ with $r_\text{s}$ being the position of the source, and 
\begin{align}
    {}_{s}\mathcal{W}_{lm\omega}&\coloneqq\Delta^{s+1}\left( {}_{s}R^\text{in}_{lm\omega}\frac{\mathrm{d}{}_{s}R^\text{up}_{lm\omega}}{\mathrm{d}r}-{}_{s}R^\text{up}_{lm\omega}\frac{\mathrm{d}{}_{s}R^\text{in}_{lm\omega}}{\mathrm{d}r} \right)\,,
\end{align}
is the Wronskian normalized by $\Delta^{s+1}$, which is constant in $r$.
We evaluate the Wronskian at $r\to\infty$, where the radial function is given by Eqs.~\eqref{eq:Rinasymp} and \eqref{eq:Rupasymp}, obtaining ${}_{-2}\mathcal{W}_{lm\omega}=2i\omega 
 C^\text{trans}_{lm\omega}B^\text{inc}_{lm\omega}$.
The solution of the radial function is given by using the radial Green function, 
\begin{align}
    {}_{s}R_{lm\omega}(r) &= -\int_{r_+}^\infty\mathrm{d}r' {}_sG_{lm\omega}(r,r') \  {}_sT_{lm\omega}(r') \nonumber \\
    &= \frac{1}{{}_{s}\mathcal{W}_{lm\omega}}\Biggl({}_{s}R^\text{up}_{lm\omega}\int_{r_+}^r\mathrm{d}r'\Delta^{s}{}_{s}R^\text{in}_{lm\omega}{}_{s}T_{lm\omega} + {}_{s}R^\text{in}_{lm\omega}\int_r^\infty\mathrm{d}r'\Delta^{s}{}_{s}R^\text{up}_{lm\omega}{}_{s}T_{lm\omega}\Biggr). 
    \label{eq:sRlmomega}
\end{align}
Finally, we obtain the solution $\Psi_4$ by substituting this radial solution~\eqref{eq:sRlmomega} into Eq.~\eqref{eq:psis}.

Hereafter, we focus on gravitational waves sourced by the motion of point particles.
The total energy-momentum tensor is then given by the sum of the energy-momentum tensor of each point particle labeled by $p$ with the mass $\mu_p$~\footnote{Eq.(5.47) in Ref.~\cite{Maggiore:2007ulw}.}, namely, 
\begin{align}
    T^{\mu\nu} &= \frac{1}{\sqrt{-g}}\sum_p\mu_p\frac{\mathrm{d}t}{\mathrm{d}\tau}\frac{\mathrm{d}x^\mu}{\mathrm{d}t}\frac{\mathrm{d}x^\nu}{\mathrm{d}t}\delta^{(3)}(x^i-x_p^i(t))\nonumber\\
    &= \frac{1}{\Sigma\sin\theta}\sum_p\mu_p\frac{\mathrm{d}t}{\mathrm{d}\tau}\frac{\mathrm{d}x^\mu}{\mathrm{d}t}\frac{\mathrm{d}x^\nu}{\mathrm{d}t}\delta(r-r_p(t))\delta(\theta-\theta_p(t))\delta(\varphi-\varphi_p(t)).
    \label{eq:EMTofpp}
\end{align}
Suppose that the null tetrad components of the energy-momentum tensor can be expressed as
\begin{subequations}
    \label{eq:TnnTnmbarTmbarmbar}
    \begin{align}
    \begin{split}
        T_{nn} &= \sum_p\mu_p\frac{C_{nn}}{\sin\theta}\delta(r-r_p(t))\delta(\theta-\theta_p(t))\delta(\varphi-\varphi_p(t)),
    \end{split}\\
    \begin{split}
        T_{n\bar{m}} &= \sum_p\mu_p\frac{C_{n\bar{m}}}{\sin\theta}\delta(r-r_p(t))\delta(\theta-\theta_p(t))\delta(\varphi-\varphi_p(t)),
    \end{split}\\
    \begin{split}
        T_{\bar{m}\bar{m}} &= \sum_p\mu_p\frac{C_{\bar{m}\bar{m}}}{\sin\theta}\delta(r-r_p(t))\delta(\theta-\theta_p(t))\delta(\varphi-\varphi_p(t)),
    \end{split}
    \end{align}
\end{subequations}
and the sources are bounded in a finite $r$, the source term~\eqref{eq:sTlmomega} can be rewritten as
\begin{align}
    {}_{s}T_{lm\omega} = \sum_p\mu_p\int_{-\infty}^\infty\mathrm{d}t e^{i\omega t -im\varphi_p(t)}\Delta^2\biggl[&(A_{nn0}+A_{n\bar{m}0}+A_{\bar{m}\bar{m}0})\delta(r-r_p(t)) \nonumber\\
    &+ \partial_r\bigl((A_{n\bar{m}1}+A_{\bar{m}\bar{m}1})\delta(r-r_p(t))\bigr) + \partial_r\partial_r\bigl((A_{\bar{m}\bar{m}2})\delta(r-r_p(t))\bigr)\biggr],
    \label{eq:sTlmomegaandA}
\end{align}
where
\begin{subequations}
    \label{eq:sourceA}
    \begin{align}
    \begin{split}
        A_{nn0} &= \frac{-2}{\sqrt{2\pi}\Delta^2}\rho^{-2}\bar{\rho}^{-1}C_{nn}\hat{L}_1^\dagger\bigl(\rho^{-4}\hat{L}_2^\dagger(\rho^3 \ {}_{-2}S_{lm\omega})\bigr),
    \end{split}\\
    \begin{split}
        A_{n\bar{m}0} &= \frac{2}{\sqrt{\pi}\Delta}\rho^{-3}C_{n\bar{m}}\Biggl[\Bigl(\hat{L}_2^\dagger \ {}_{-2}S_{lm\omega}\Bigr)\Biggl(\frac{iK}{\Delta}+\rho+\bar{\rho}\Biggr)-a\sin\theta \ {}_{-2}S_{lm\omega}\frac{K}{\Delta}(\bar{\rho}-\rho)\Biggr],
    \end{split}\\
    \begin{split}
        A_{\bar{m}\bar{m}0} &= \frac{-1}{\sqrt{2\pi}}\rho^{-3}\bar{\rho}C_{\bar{m}\bar{m}} \ {}_{-2}S_{lm\omega}\Biggl[-i\partial_r\frac{K}{\Delta}-\frac{K^2}{\Delta^2}+2i\rho\frac{K}{\Delta}\Biggr],
    \end{split}\\
    \begin{split}
        A_{n\bar{m}1} &= \frac{2}{\sqrt{\pi}\Delta}\rho^{-3}C_{n\bar{m}}\biggl[\hat{L}_2^\dagger \ {}_{-2}S_{lm\omega} + ia\sin\theta(\bar{\rho}-\rho){}_{-2}S_{lm\omega}\biggr],
    \end{split}\\
    \begin{split}
        A_{\bar{m}\bar{m}1} &= \frac{-2}{\sqrt{2\pi}}\rho^{-3}\bar{\rho}C_{\bar{m}\bar{m}} \ {}_{-2}S_{lm\omega}\biggl(\frac{iK}{\Delta}+\rho\biggr),
    \end{split}\\
    \begin{split}
        A_{\bar{m}\bar{m}2} &= \frac{-1}{\sqrt{2\pi}}\rho^{-3}\bar{\rho}C_{\bar{m}\bar{m}}{}_{-2}S_{lm\omega}.
    \end{split}
    \end{align}
\end{subequations}
Note that Eqs.~\eqref{eq:sourceA} contain $\theta_p$ as the argument.
Substituting Eq.~\eqref{eq:sTlmomegaandA} into Eq.~\eqref{eq:sRlmomega}, we have
\begin{align}
    {}_{s}R_{lm\omega}(r;\theta_p)
    &=
     \left\{ \,
        \begin{aligned}
           &\frac{1}{{}_{s}\mathcal{W}_{lm\omega}}{}_{s}R^\text{up}_{lm\omega}(r)\sum_p\mu_p\int_{-\infty}^{\infty}\mathrm{d}t e^{i\omega t-im\varphi_p(t)} {}_{s}W^\text{in}_{lm\omega}(\theta_p) & \text{for } & r_\text{s} < r,\\
            &\frac{1}{{}_{s}\mathcal{W}_{lm\omega}}{}_{s}R^\text{in}_{lm\omega}(r)\sum_p\mu_p\int_{-\infty}^{\infty}\mathrm{d}t e^{i\omega t-im\varphi_p(t)} {}_{s}W^\text{up}_{lm\omega}(\theta_p) & \text{for } & r < r_\text{s},
        \end{aligned}
    \right.
    \label{eq:sRlmomegaW}
\end{align}
where
\begin{align}
    {}_{s}W_{lm\omega}^\text{(in/up)}(\theta_p) = \biggl[ \Bigl(A_{nn0}+A_{n\bar{m}0}+A_{\bar{m}\bar{m}0}\Bigr){}_{s}R_{lm\omega}^\text{(in/up)} - \Bigl(A_{n\bar{m}1}+A_{\bar{m}\bar{m}1}\Bigr)\frac{\mathrm{d}{}_{s}R_{lm\omega}^\text{(in/up)}}{\mathrm{d}r} +\Bigl(A_{\bar{m}\bar{m}2}\Bigr)\frac{\mathrm{d}^2 {}_{s}R_{lm\omega}^\text{(in/up)}}{\mathrm{d}r^2}  \biggr]_{r=r_p(t)}.
\end{align}

\section{\label{sec:lensing}Gravitational waves lensed by a Kerr black hole:\\ on-axis and aligned rotation axis case}
\begin{figure}[h]
    \includegraphics[width=0.5\columnwidth]{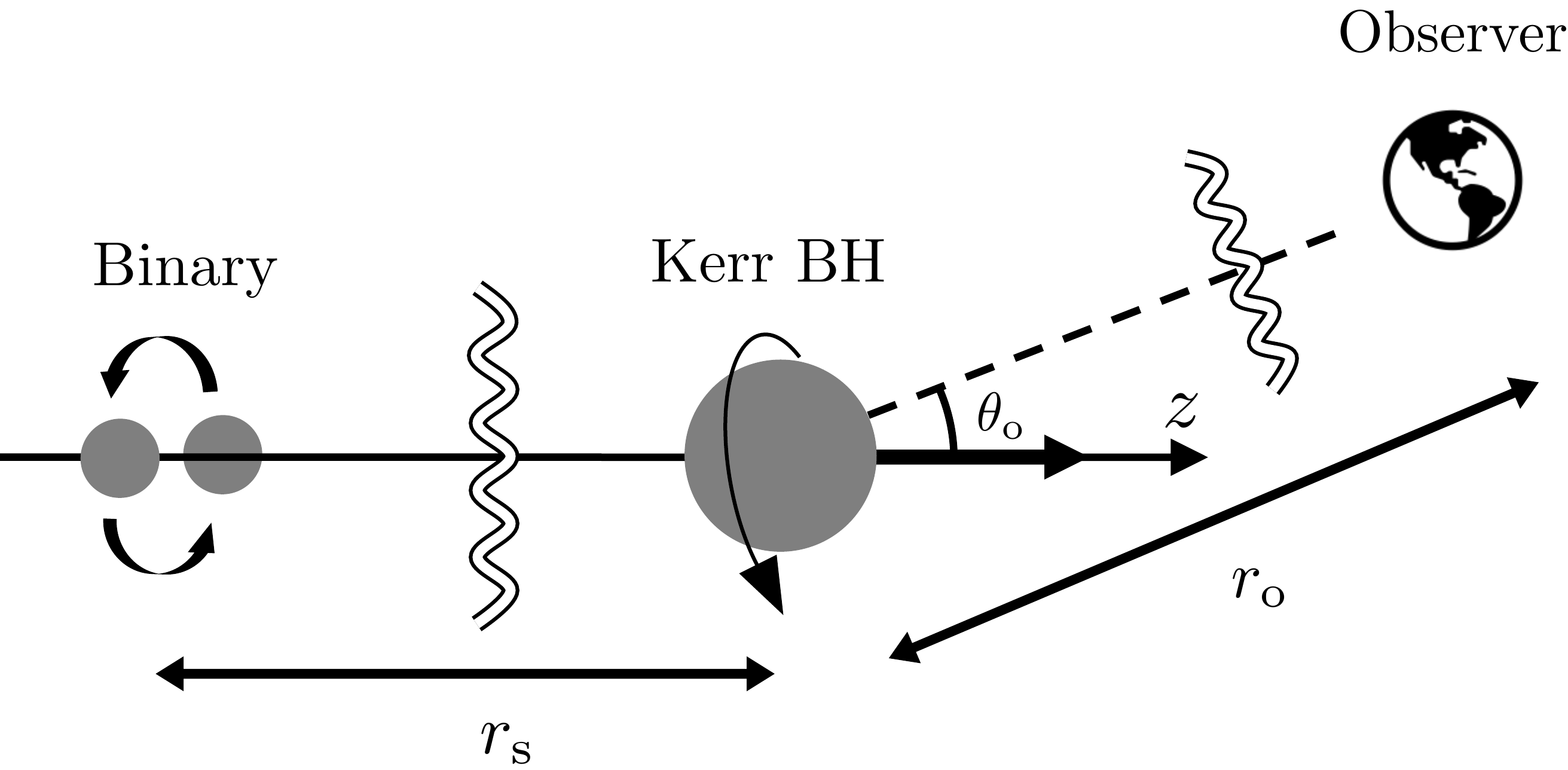}
    \caption{\label{fig:config}The configuration of the gravitational lensing.}
\end{figure}

In this section, we investigate the gravitational waves lensed by a Kerr black hole assuming the gravitational waves are sourced by an equal mass circular binary. 
We consider a situation where the binary is on $z$-axis, which is the rotation axis of the Kerr black hole at the origin, and the rotation axis of the binary orbit is aligned with $z$-axis (see Fig.~\ref{fig:config}).
We assume that the Kerr black hole does not affect the motion of the binary system.
We denote the equal mass of the stars as $\mu$.
The positions of the binary stars $x_1^\mu$ and $x_2^\mu$ are specified in spherical coordinates as 
\begin{align}
    r_1(t) = r_2(t) =\sqrt{r_\text{s}^2+d^2} \simeq r_\text{s},\quad 
    \theta_1(t) = \theta_2(t) = \cos^{-1}\biggl(\frac{-r_\text{s}}{\sqrt{r_\text{s}^2+d^2}}\biggr),\quad  
    \varphi_1(t) = \omega_\text{s}t,\quad  
    \varphi_2(t) = \omega_\text{s}t + \pi,
    \label{eq:position}
\end{align}
where $r_\text{s}$ is the distance from the Kerr black hole to the center of the binary, $d$ is the radius of the binary orbit, and $\omega_\text{s}$ is the frequency of the orbital motion.
The velocities are given by
\begin{align}
    \frac{\mathrm{d}x^\mu_1}{\mathrm{d}t}=\frac{\mathrm{d}x^\mu_2}{\mathrm{d}t}=\begin{pmatrix} 1, 0, 0, \omega_\text{s} \end{pmatrix}.
    \label{eq:velocity}
\end{align}
We assume that the binary system moves in the Keplerian-like motion where the angle of the binary orbit is associated with the frequency as
\begin{align}
    (r_\text{s}\sin\theta_\text{s})^3\omega_\text{s}^2 = \frac{\mu}{4},
    \label{eq:Kepler}
\end{align}
where $r_\text{o}$ and $\theta_\text{o}$ specify the position of the observer.

Before investigating lensed gravitational waves, let us consider the polarization state of gravitational waves radiated by a binary system on the Minkowski background spacetime.
The metric in the region far away from the binary system, i.e., $r'\omega \gg 1$, is given by 
\begin{align}
    \mathrm{d}s^2=-\mathrm{d}t^2+\mathrm{d}r'^2+(1+h'_+)(r'\mathrm{d}\theta')^2+2h'_\times (r'\mathrm{d}\theta')(r'\sin\theta'\mathrm{d}\varphi')+(1-h'_+)(r'\sin\theta'\mathrm{d}\varphi')^2
    \label{eq:metricwgw}
\end{align}
where $(r',\theta',\varphi')$ is a spherical coordinate system centered at the center of the binary system, 
$h'_+$ and $h'_\times$ are the gravitational waves emitted by a quadrupole moment of the circular orbit binary system~\cite{Nakamura:1998} (see e.g., Refs.~\cite{Creighton:2011zz,Maggiore:2007ulw}) given by
\begin{align}
    \begin{aligned}
        h'_{+} &= -\frac{8\mu\omega_\text{s}^2d^2}{r'}\frac{1+\cos^2\theta'}{2}\cos(2\omega_\text{s}(t-r')-2\varphi'),\\
        h'_{\times} &= -\frac{8\mu\omega_\text{s}^2d^2}{r'}\cos\theta'\sin(2\omega_\text{s}(t-r')-2\varphi').
    \end{aligned}
    \label{eq:quadrapoleradiation}
\end{align}
For $\theta'=0$ and $\omega_\text{s}>0$, right-handed circular polarized gravitational waves are emitted since the plus mode has the same amplitude as that of the cross mode and the phase of the plus mode advances by $\pi/2$ relative to that of the cross mode.
Similarly, for $\theta'=0$ and $\omega_\text{s}<0$, the emitted gravitational waves exhibit left-handed circular polarization, since the phase of the plus mode is delayed by $\pi/2$ relative to that of the plus mode.
Therefore, the situation we consider corresponds to the gravitational lensing of right-handed and left-handed circular polarized gravitational waves for $\omega_\text{s}>0$ and $\omega_\text{s}<0$, respectively.
Throughout this paper, we consider only positive $\omega_s$ without loss of generality.

We calculate the Weyl scalar $\Psi_4$ following the Teukolsky formalism.
Substituting Eqs.~\eqref{eq:velocity} into Eq.\eqref{eq:EMTofpp}, we have
\begin{subequations}
    \label{eq:CnnCnmbarCmbarmbar}
    \begin{align}
    \begin{split}
        C_{nn} &= \frac{1}{\Sigma}\frac{\mathrm{d}t}{\mathrm{d}\tau}\left(\frac{\mathrm{d}x^t}{\mathrm{d}t}n_t + \frac{\mathrm{d}x^\varphi}{\mathrm{d}t}n_\varphi\right)^2 
        = \frac{1}{\Sigma}\frac{\mathrm{d}t}{\mathrm{d}\tau}\left(\frac{\Delta}{2\Sigma}\right)^2\left(-1+a\omega_\text{s}\sin^2\theta\right)^2,
    \end{split}\\
    \begin{split}
        C_{n\bar{m}} &= \frac{1}{\Sigma}\frac{\mathrm{d}t}{\mathrm{d}\tau}\left(\frac{\mathrm{d}x^t}{\mathrm{d}t}n_t + \frac{\mathrm{d}x^\varphi}{\mathrm{d}t}n_\varphi\right)\left(\frac{\mathrm{d}x^t}{\mathrm{d}t}\bar{m}_t + \frac{\mathrm{d}x^\varphi}{\mathrm{d}t}\bar{m}_\varphi\right) \\
         &=  \frac{1}{\Sigma}\frac{\mathrm{d}t}{\mathrm{d}\tau}\left(\frac{\Delta}{2\Sigma}\right)\left(\frac{i\rho\sin\theta}{\sqrt{2}}\right)\left(-1+a\omega_\text{s}\sin^2\theta\right)\left(a-\omega_\text{s}(r^2+a^2)\right),
    \end{split}\\
    \begin{split}
        C_{\bar{m}\bar{m}} &= \frac{1}{\Sigma}\frac{\mathrm{d}t}{\mathrm{d}\tau}\left(\frac{\mathrm{d}x^t}{\mathrm{d}t}\bar{m}_t + \frac{\mathrm{d}x^\varphi}{\mathrm{d}t}\bar{m}_\varphi\right)^2 = \frac{1}{\Sigma}\frac{\mathrm{d}t}{\mathrm{d}\tau}\left(\frac{i\rho\sin\theta}{\sqrt{2}}\right)^2\left(a-\omega_\text{s}(r^2+a^2)\right)^2.
    \end{split}
    \end{align}
\end{subequations}
The prefactor $\frac{\mathrm{d}t}{\mathrm{d}\tau}$ is calculated from the definition of proper time $g_{\mu\nu}\frac{\mathrm{d}x^\mu}{\mathrm{d}\tau}\frac{\mathrm{d}x^\nu}{\mathrm{d}\tau} = -1$.
By using Eq.~\eqref{eq:velocity} and Eq.~\eqref{eq:metric}, we have
\begin{align}
    \frac{\mathrm{d}t}{\mathrm{d}\tau} = \frac{1}{ \sqrt{-g_{tt} - 2\omega_\text{s}g_{t\varphi} - \omega_\text{s}^2 g_{\varphi\varphi} } }.
    \label{eq:dtdtau}
\end{align}
Putting Eq.~\eqref{eq:CnnCnmbarCmbarmbar} and Eq.~\eqref{eq:dtdtau} into Eqs.~\eqref{eq:sourceA}, we obtain the source term.
Then, we can calculate the inhomogeneous radial function from Eq.~\eqref{eq:sRlmomegaW}.
The time integral in Eq.~\eqref{eq:sRlmomegaW} becomes the delta function as
\begin{align}
    \sum_p\mu_p\int_{-\infty}^{\infty}\mathrm{d}t e^{i\omega t-im\varphi_p(t)} {}_{s}W_{lm\omega}=&
    \left\{ \,
        \begin{aligned}
            & 4\pi\mu\delta(\omega-m\omega_\text{s})\ {}_{s}W_{lm\omega} & \text{for } & \text{even }m,\\
            & 0 & \text{for } & \text{odd }m.
        \end{aligned}
    \right.
    \label{eq:timeintegral}
\end{align}
We have used the fact that, for odd $m$, the anti-phase oscillations of $\varphi_1$ and $\varphi_2$ lead to the cancellation in the summation over the sources.
Finally, we can obtain the Weyl scalar by substituting the inhomogeneous radial function into Eq.~\eqref{eq:psis}.

\begin{figure}[htbp]
    \centering
    \includegraphics[keepaspectratio, width=0.6\columnwidth]{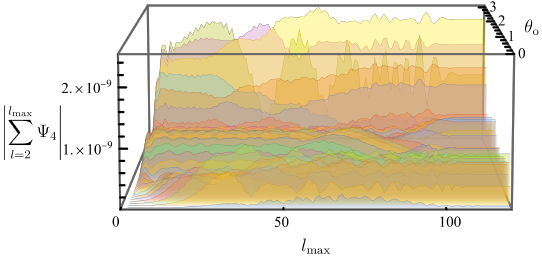}
    \captionsetup{justification = raggedright, singlelinecheck = false}
    \caption{The convergence of the summation over $l$ for $m=-2$. The binary with the orbital frequency $M\omega_\text{s}=1/2$ is located at $r_\text{s} = 100M $. The observer is located at $(r_\text{o}, \varphi_\text{o})=(200M, 0)$. The Kerr parameter is $a=99/100$. The stellar mass of the binary is $\mu=10^{-4}M.$ The summation over $l$ converges around $l_\text{max} = r_{>} \omega=100$, where we truncate the summation.}
    \label{fig:convergence}
\end{figure}

The summation over $l$ in Eq.~\eqref{eq:psis} can be truncated around $l \sim r_{<} \omega$ for the following reason.
The potential of the Schr\"odinger form of the radial Teukolsky equation~\eqref{eq:potentialatlarger} becomes $V(r) \sim \omega^2 - l^2 / r^2$ for $r \to \infty$ and large $l$, where we have used $\lambda \sim l^2 $.
For $l > r_{<} \omega $, the potential at the source or the observer is positive, and thus the Green function is exponentially suppressed.
Therefore, we only need to sum up to $l \sim r_{<} \omega $.
Fig.~\ref{fig:convergence} shows $\Psi_4$ for $m=-2$ summed up to $l_\text{max}$.
Indeed, the summation converges around $l_\text{max} \sim r_{<} \omega = 100$.

The summation $\sum_{m=-\infty}^{m=\infty}$ in Eq.~\eqref{eq:psis} can be replaced by the summation $\sum_{m=\pm2}$ for the following two reasons.
The first reason is that $m=\pm 2$ modes are basically dominant.
The functions ${}_{-2}S_{lm\omega}$, $\hat{L}_2^\dagger \ {}_{-2}S_{lm\omega}$, and $\hat{L}_1^\dagger\hat{L}_2^\dagger \ {}_{-2}S_{lm\omega}$ in the source term have non-zero values at $\theta=\pi$ only for $m=-2$, $-1$, and $0$,\footnote{
For $\theta=0,\pi$, we have some useful relations~\cite{Sasaki:1981sx},
\begin{align}
    \Bigl[\hat{L}_1^\dagger\hat{L}_2^\dagger\ {}_{-2}S_{lm\omega}\Bigr]_{\theta=0,m=0}&=\biggl[8\frac{{}_{-2}S_{lm\omega}}{\sin^2\theta}\biggr]_{\theta=0,m=0},\\
    \Bigl[\hat{L}_2^\dagger\ {}_{-2}S_{lm\omega}\Bigr]_{\theta=0,m=1}&=\biggl[2\frac{{}_{-2}S_{lm\omega}}{\sin\theta}\biggr]_{\theta=0,m=1},\\
    \Bigl[\hat{L}_1^\dagger\hat{L}_2^\dagger\ {}_{-2}S_{lm\omega}\Bigr]_{\theta=\pi,m=0}&=\biggl[8\frac{{}_{-2}S_{lm\omega}}{\sin^2\theta}\biggr]_{\theta=\pi,m=0},\\
    \Bigl[\hat{L}_2^\dagger\ {}_{-2}S_{lm\omega}\Bigr]_{\theta=\pi,m=-1}&=\biggl[-2\frac{{}_{-2}S_{lm\omega}}{\sin\theta}\biggr]_{\theta=\pi,m=-1}.
\end{align}
One can verify these relations by substituting the expression expanded around $\theta=0$ (see e.g., \cite{Casals:2018cgx}),
\begin{align}
    {}_{s}S_{lm\omega}(\theta)=\theta^{|m+s|}\sum_{k=0}^\infty a_k \theta^k,
\end{align}
or the expression expanded around $\theta=\pi$,
\begin{align}
    {}_{s}S_{lm\omega}(\theta)=(\pi-\theta)^{|m-s|}\sum_{k=0}^\infty b_k (\pi-\theta)^k.
\end{align}
} respectively, i.e.,
\begin{align}
    {}_{-2}S_{lm\omega}(\pi) =& \biggl[{}_{-2}S_{l(-2)\omega}(\theta)\biggr]_{\theta=\pi}\delta_{m(-2)},\\
    \hat{L}_2^\dagger {}_{-2}S_{lm\omega}(\pi) =& \biggl[-2\frac{{}_{-2}S_{l(-1)\omega}(\theta)}{\sin\theta}\biggr]_{\theta=\pi}\delta_{m(-1)},\\
    \hat{L}_1^\dagger \hat{L}_2^\dagger {}_{-2}S_{lm\omega}(\pi) =& \biggl[8\frac{{}_{-2}S_{l0\omega}(\theta)}{\sin^2\theta}\biggr]_{\theta=\pi}\delta_{m0}.
\end{align}
Since the polar angle coordinates of the source $\theta_\text{1}$ and $\theta_\text{2}$ are only slightly deviated from $\pi$, 
$m=-2$, $-1$, and $0$ modes are dominant and the other modes are suppressed.
Using Eq.~\eqref{eq:timeintegral}, we can find that $m=-1$ mode vanishes and $m=0$ mode does not oscillate.
Thus, only $m=-2$ mode is worth investigating.
However, when considering gravitational waves in the forward direction $\theta_\text{o}\sim 0$, $m=2$ mode is dominant rather than $m=-2$ modes since the angular function in Eq.~\eqref{eq:psis} is ${}_{-2}S_{lm\omega}(\theta) \propto \theta^{|m+s|}$ and $m=-2$ modes are suppressed.
This suppression is shown in Fig.~\ref{fig:rs10} which will appear later.
The second reason is our knowledge that the frequency of gravitational waves sourced by quadrupole moment~\eqref{eq:quadrapoleradiation} is twice the orbital frequency of the source.

\subsection{Case of source and observer close to Kerr black hole lens}
We consider the gravitational lensing without using geometrical optics and large-distance asymptotic approximations for the case where the source and observer are close to the Kerr black hole lens.
We use the package {\tt Teukolsky} and the package {\tt SpinWeightedSpheroidalHarmonics} in the Black Hole Perturbation Toolkit~\cite{BHPToolkit} to calculate the radial and angular functions, respectably.

\subsubsection{\label{sec:numericalresults}Numerical results}
\begin{figure}[htbp]
    \begin{tabular}{cc}
        \begin{minipage}[b]{0.45\linewidth}
            \centering
            \includegraphics[keepaspectratio, width=\columnwidth]{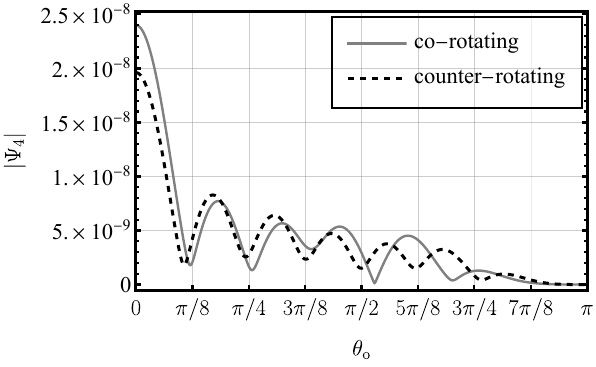}
            \subcaption{$m=2$}
        \end{minipage} &
        \begin{minipage}[b]{0.45\linewidth}
            \centering
            \includegraphics[keepaspectratio, width=\columnwidth]{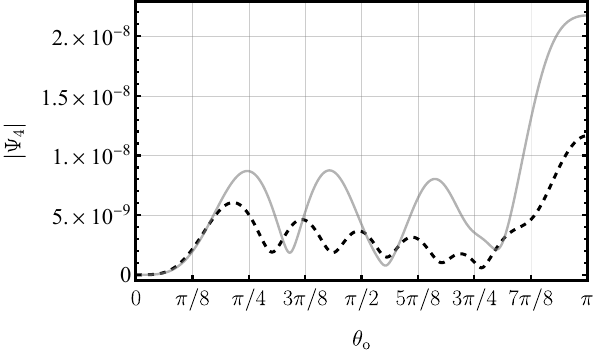}
            \subcaption{$m=-2$}
        \end{minipage} \\ 
        \begin{minipage}[b]{0.45\linewidth}
            \centering
            \includegraphics[keepaspectratio, width=\columnwidth]{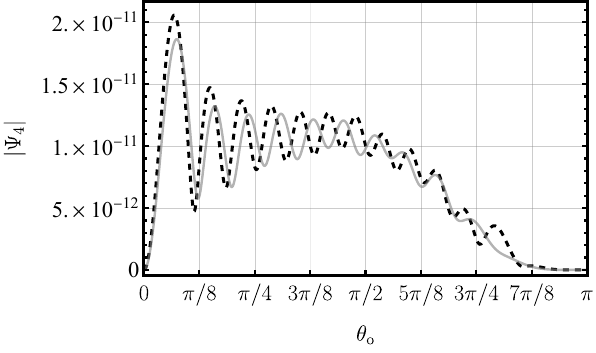}
            \subcaption{$m=4$}
        \end{minipage} &
        \begin{minipage}[b]{0.45\linewidth}
            \centering
            \includegraphics[keepaspectratio, width=\columnwidth]{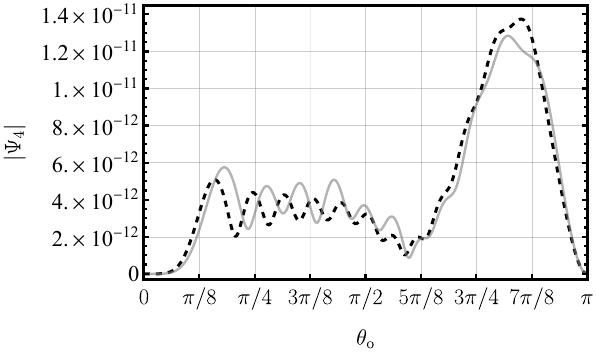}
            \subcaption{$m=-4$}
        \end{minipage}
    \end{tabular}
    \captionsetup{justification = raggedright, singlelinecheck = false}
    \caption{The absolute value of $\Psi_4$ sourced by a binary system with $\omega_\text{s}=1/2$ at $r_\text{s}=10M$. The observer is located at $(r_\text{o}, \varphi_\text{o})=(20M, 0)$. The Kerr parameter is $a/M=99/100$ (gray line) and $a/M=-99/100$ (black dashed line). The stellar mass of the binary is $\mu=10^{-4}M.$ We truncate the summation over $l$ around $l=40$.}
    \label{fig:rs10}
\end{figure}

Fig.~\ref{fig:rs10} shows the angular dependence of the absolute value of the Weyl scalar source by co-rotating (gray line) and counter-rotating (black dashed line) source at $r_\text{s}=10M$.
In the co-rotating case, the angular momentum vector of the Kerr black hole is aligned with that of the binary system, whereas in the counter-rotating case, the angular momentum vector is anti-aligned with that of the binary system.
We find the difference between the co-rotating case and counter-rotating case, which reflects the distinction in the propagation of the left- and right-handed gravitational waves.
As mentioned above, the Weyl scalar for $m=\pm 4$ is smaller than that for $m=\pm 2$.

\begin{figure}[htbp]
    \begin{minipage}[b]{0.45\linewidth}
        \centering
        \includegraphics[keepaspectratio, width=\columnwidth]{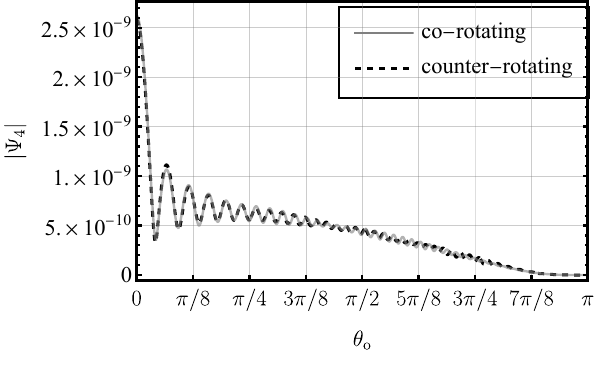}
        \subcaption{$m=2$}
    \end{minipage}
    \begin{minipage}[b]{0.45\linewidth}
        \centering
        \includegraphics[keepaspectratio, width=\columnwidth]{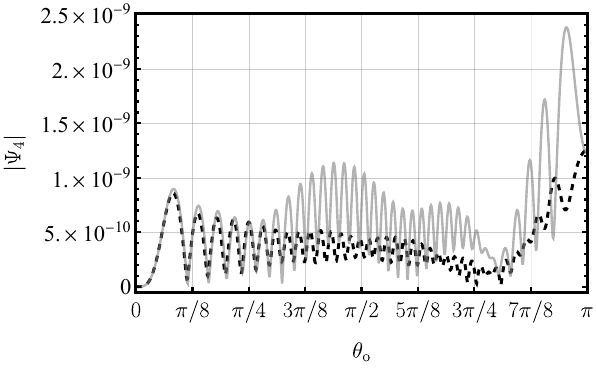}
        \subcaption{$m=-2$}
    \end{minipage}
    \captionsetup{justification = raggedright, singlelinecheck = false}
    \caption{The absolute value of $\Psi_4$ sourced by a binary system with $\omega_\text{s}=1/2$ at $r_\text{s} = 100M $. The observer is located at $(r_\text{o}, \varphi_\text{o})=(200M, 0)$. The Kerr parameter is $a/M=99/100$ (gray line) and $a/M=-99/100$ (black dashed line). The stellar mass of the binary is $\mu=10^{-4}M.$ We truncate the summation up to $l=120$.}
    \label{fig:rs100}
\end{figure}

\begin{figure}[htbp]
    \begin{minipage}[b]{0.45\linewidth}
        \centering
        \href{https://zenodo.org/records/13141200/preview/scattering_gw_co-rotating_omegas05.mp4?include_deleted=0}{\includegraphics[keepaspectratio, width=\columnwidth]{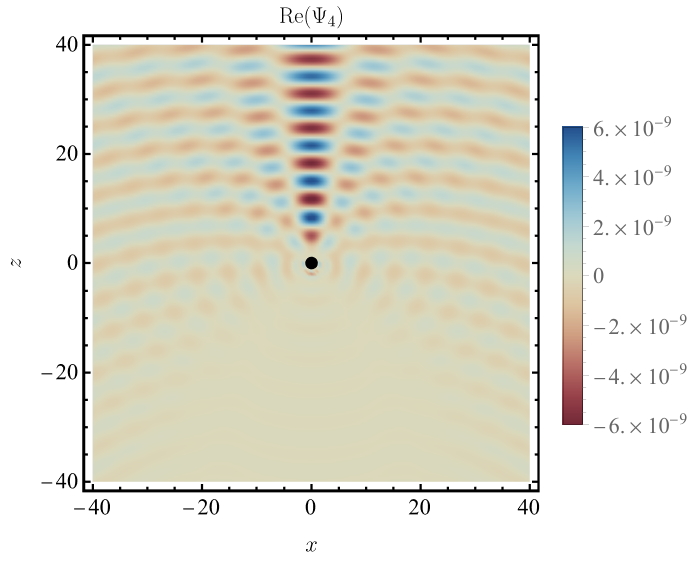}}
        \subcaption{co-rotating $a/M=99/100$}
    \end{minipage}
    \begin{minipage}[b]{0.45\linewidth}
        \centering
        \href{https://zenodo.org/records/13141200/preview/scattering_gw_counter-rotating_omegas05.mp4?include_deleted=0}{\includegraphics[keepaspectratio, width=\columnwidth]{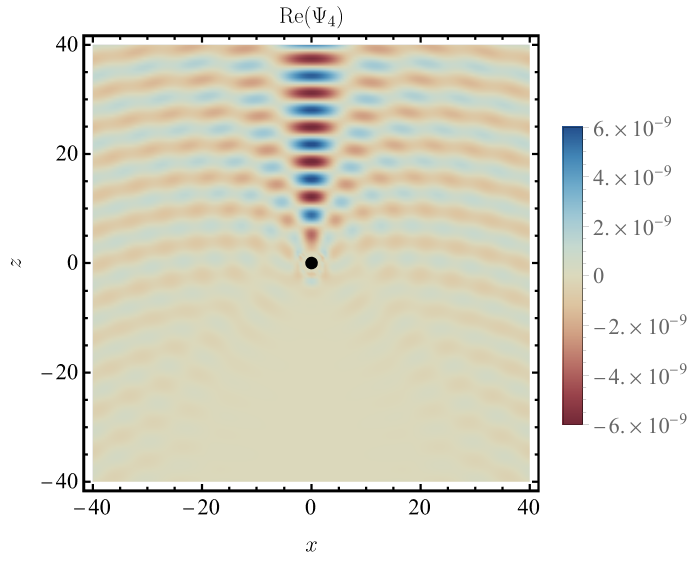}}
        \subcaption{counter-rotating $a/M=-99/100$}
    \end{minipage}
    \captionsetup{justification = raggedright, singlelinecheck = false}
    \setfloatlink{https://zenodo.org/records/11906041}
    \caption{The density plot of $\mathrm{Re}(\Psi_4)$ for the source with $M\omega_\text{s}=1/2$ at $(r_\text{s},\theta_\text{s})=(100M,\pi)$. The Kerr parameter is $a/M=99/100$ (left) and $a/M=-99/100$ (right). The mass of the binary is $\mu=10^{-4}M$. We take the sum of $m=-2$ and $m=2$ in the summation over $m$. The summation over $l$ is truncated up to $l=60$. We add an animation link to these images.}
    \label{fig:scattering}
\end{figure}

Fig.~\ref{fig:rs100} and Fig.~\ref{fig:scattering} show the angular dependence and the density for $r_\text{s}=100M$ and $r_\text{o}=200M$, respectively.
Compared to the previous case for $r_\text{s}=10M$ and $r_\text{o}=20M$, the difference between the co-rotating case and counter-rotating case in the forward direction $\theta_\text{o}\sim 0$ is small, but it is still present in the backward direction $\theta_\text{o}\sim \pi$.
The enhancement of the spin effect is attributed to the fact that the low-$l$ modes dominantly contribute, which are sensitive to the spin because they pass near the black hole.
However, at exactly $\theta_\text{o}=\pi$, the difference is small.
This means that only the gravitational waves directly propagating from the source to the observer reach the observer, while the gravitational waves reflected by the lens do not.
This result is consistent with the fact that the differential scattering cross-section of plane gravitational waves is absent for $\theta_\text{o}=\pi$ when superradiance~\cite{Starobinsky:1973aij,Starobinsky:1974,Dolan:2008kf} does not occur.
We can see only the outgoing waves in Fig.~\ref{fig:scattering}.
This is because the ingoing wave is suppressed (see Eqs.~\eqref{eq:Rinasymp} and \eqref{eq:sRlmomegaW}).

\begin{figure}[htbp]
    \begin{minipage}[b]{0.45\linewidth}
        \centering
        \includegraphics[keepaspectratio, width=\columnwidth]{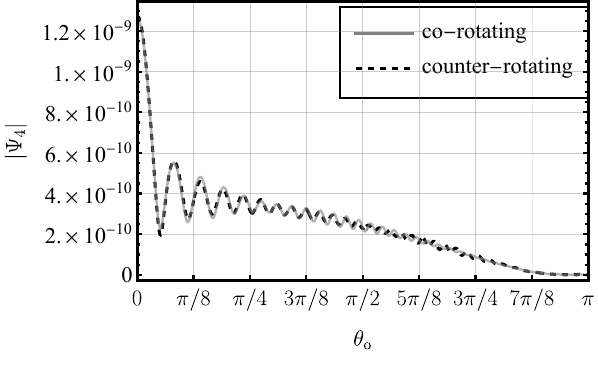}
        \subcaption{$m=2$}
    \end{minipage}
    \begin{minipage}[b]{0.45\linewidth}
        \centering
        \includegraphics[keepaspectratio, width=\columnwidth]{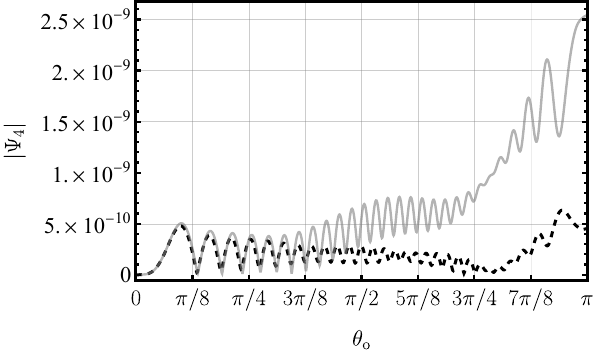}
        \subcaption{$m=-2$}
    \end{minipage}
    \captionsetup{justification = raggedright, singlelinecheck = false}
    \caption{The absolute value of $\Psi_4$ for $|\omega_\text{s}|=0.4$. The other parameters are the same as Fig.\ref{fig:rs100}. One can find the amplification in the backward direction due to the superradiance phenomenon.}
    \label{fig:rs100_superradiance}
\end{figure}

\begin{figure}[htbp]
    \begin{minipage}[b]{0.45\linewidth}
        \centering
        \href{https://zenodo.org/records/13141200/preview/scattering_gw_co-rotating_omegas04.mp4?include_deleted=0}{\includegraphics[keepaspectratio, width=\columnwidth]{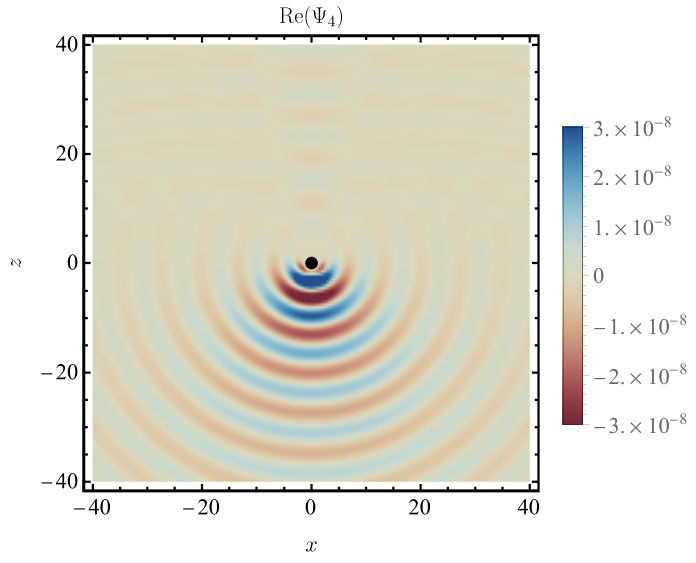}}
        \subcaption{co-rotating $a/M=99/100$}
    \end{minipage}
    \begin{minipage}[b]{0.45\linewidth}
        \centering
        \href{https://zenodo.org/records/13141200/preview/scattering_gw_counter-rotating_omegas04.mp4?include_deleted=0}{\includegraphics[keepaspectratio, width=\columnwidth]{./figure/scattering_gw_counter-rotating_omegas05.pdf}}
        \subcaption{counter-rotating $a/M=-99/100$}
    \end{minipage}
    \captionsetup{justification = raggedright, singlelinecheck = false}
    \caption{The density plot of $\mathrm{Re}(\Psi_4)$ when the superradiance occur. The summation over $l$ is truncated up to $l=50$. The frequency of the orbital motion is $M\omega_\text{s}=4/10$. The other parameters are the same as Fig.~\ref{fig:scattering}. We add an animation link to these images.}
    \label{fig:scattering_superradiance}
\end{figure}

The superradiance occurs for $\omega < m a / (2Mr_+)$.
Fig.~\ref{fig:rs100_superradiance} and Fig.~\ref{fig:scattering_superradiance} show the angular dependence and the density for $M|\omega_\text{s}|=0.4$ when the supperradiance occurs.
One can find $\Psi_4$ in the backward direction for the co-rotating case is amplified with respect to that for the counter-rotating case.

\subsubsection{\label{sec:amplification}Amplification factor at $\theta=0$}
The amplification factor is typically formulated in the frequency domain. However, in this paper, we focus on monochromatic waves, making a Fourier transform unnecessary. To obtain the amplification factor, we need the unlensed Weyl scalar $\Psi_4^\text{ul}$, i.e., Weyl scalar for the case of no lens.
Performing coordinate transformation from $(t,r',\theta',\varphi')$ to $(t,r,\theta,\varphi)$, the metric for $\theta\simeq 0$ given in Eq.~\eqref{eq:metricwgw} becomes 
\begin{align}
    \mathrm{d}s^2=-\mathrm{d}t^2+\mathrm{d}r^2+(1+h'_+)(r\mathrm{d}\theta)^2+2h'_\times (r\mathrm{d}\theta)(r\sin\theta\mathrm{d}\varphi)+(1-h'_+)(r\sin\theta\mathrm{d}\varphi)^2,
    \label{eq:hphc2}
\end{align}
where $h'_+$ and $h'_\times$ are given in Eq.~\eqref{eq:quadrapoleradiation}.
At the leading order of the geometrical optics and large-distance approximation where $r'\omega \gg 1$, the metric~\eqref{eq:hphc2} yields unlensed Weyl scalar as (see e.g., Ref.~\cite{Nakamura:1987zz})
\begin{align}
    \Psi_4^\text{ul} &\simeq \frac{1}{8}\left(\partial_t - \partial_r\right)^2(h'_+-ih'_\times) \nonumber \\
    &= -\frac{16\mu\omega_\text{s}^4d^2}{r'}\exp(-2i\omega_\text{s}(t-r')+2i\varphi).
    \label{eq:unlensedPsi4}
\end{align}
The amplification factor for $\theta=0$ is written as
\begin{align}
    F(\omega)=\frac{\Psi_4}{\Psi_4^\text{ul}}.
\end{align}
The lensed Weyl scalar $\Psi_4$ is calculated from substitution of Eq.~\eqref{eq:sRlmomega} into Eq.~\eqref{eq:psis}, and $\Psi_4^\text{ul}$ is calculated from Eq.~\eqref{eq:unlensedPsi4}.

\begin{figure}[htbp]
    \centering
    \includegraphics[keepaspectratio, width=\columnwidth]{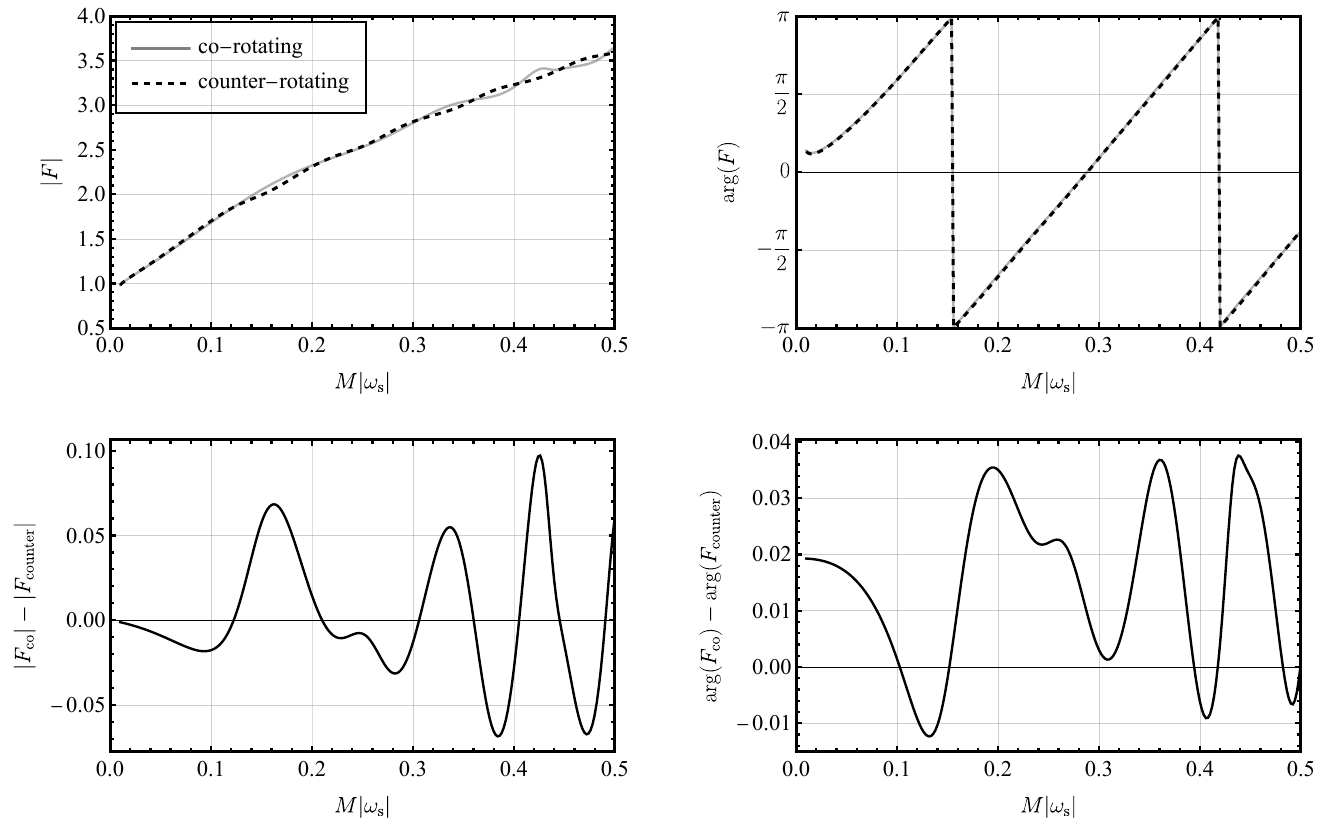}
    \captionsetup{justification = raggedright, singlelinecheck = false}
    \caption{The absolute value $|F|$ (top-left) and the phase $\mathrm{arg}(F)$ (top-right) of the amplification factor for $(m,a,\mu,r_\text{s},r_\text{o},\theta_\text{o})=(2,0.99M,10^{-4}M,100M,200M,0)$, and  the difference in the absolute value (bottom-left) and the phase (bottom-right) between the amplification factor for the co-rotating case $F_\mathrm{co}$ and that for the counter-rotating case $F_\mathrm{counter}$.}
    \label{fig:amplitifcation}
\end{figure}

Fig.~\ref{fig:amplitifcation} shows the amplification factor of co-rotating and counter-rotating cases for $t=\varphi=0$ with the frequency from $M\omega_\text{s}=10^{-3}$ to $M\omega_\text{s}=0.5$, where the wave effects are prominent.
We have ensured that the result does not change as long as we alter the precision in numerical calculation and the truncation scale $l_{\rm max}$, suggesting that the result is robust. 

We find that the absolute value $|F|$ of the amplification factor approaches unity as frequency decreases. 
This means that when we take the limit $M\omega_{\text{s}} \to 0$, the lensed $\Psi_4$ tends towards the unlensed $\Psi_4^\text{ul}$ which is obtained by analytic calculation \eqref{eq:unlensedPsi4}, i.e., the low-frequency waves are not affected by the lensing.
However, we see that the phase $\arg(F)$ in Fig.~\ref{fig:amplitifcation} does not go to $0$ at the limit $M\omega_{\text{s}} \to 0$. This is because Eq.~\eqref{eq:unlensedPsi4} does not include the correction term of $O(1/\omega r')$, getting less accurate at lower frequencies. 

On the other hand, in the high-frequency regime, $|F|$ oscillates with a small amplitude.
Oscillations similar to this have been found for the scattering of scalar waves by the Schwarzschild black hole~\cite{Nambu:2019sqn,Motohashi:2021zyv}.
They concluded that this oscillation results from the diffraction between the direct ray that propagates far from the black hole and the winding ray which encircles the black hole.
The period of this oscillation is approximately $M\Delta\omega = M2\Delta \omega_\text{s} \simeq 0.2$ around $M\omega_\text{s}=0.4$, consistent with the result of the present paper.

We observe a monotonic increase in the phase $\arg(F)$ in Fig.~\ref{fig:amplitifcation}.
We confirm that the behavior of the monotonic increase of $\arg(F)$ does not depend on the angular momentum of the black hole and does not vary significantly with distance.
From Fig.~\ref{fig:amplitifcation}, we can read off that the monotonic increase can be approximately expressed as $\arg(F) \simeq 24 M\omega_\text{s}$.
On the other hand, from Eq.~\eqref{eq:unlensedPsi4}, the phase of the unlensed Weyl scalar increases monotonically as $\arg(\Psi_4^\text{ul}) = -2\omega_\text{s}(t-r')=600M\omega_\text{s}$. 
Since $\arg(F) = \arg(\Psi_4) - \arg(\Psi_4^\text{ul})$, we can estimate $\arg(\Psi_4)\simeq 624M\omega_\text{s}$, i.e., the phase of Weyl scalar is shifted by about $4\%$ after lensing.
Based on $\arg(\Psi_4^\text{ul}) = -2\omega_\text{s}(t-r')$ in the geometric optics approximation, it would be natural to interpret the above phase shift as a result of the difference in the path length between the lensed and unlensed paths. 

We also find a frequency-dependent differences in $\arg(F)$ between co-rotating and counter-rotating cases.
Given that its period matches the characteristic period of the interference between the direct ray and the winding ray, we expect that this phase difference is related to this interference.
The difference between the amplification factor for positive and negative spin parameters in a fixed orbital frequency can be regarded as the difference between the amplification factor for positive and negative orbital frequency for a fixed spin parameter.
In this perspective, the phase difference can be interpreted as the difference between the phase for the left- and right-handed gravitational waves.
The phase difference induces the rotation of the major axis of the polarization.

The rotation of the polarization dragged by a rotating object has long been known as the so-called ``gravitational Faraday rotation''~\cite{Dehnen:1973xa,CONNORS:1977,Connors:1980,Piran:1985,Chandrasekhar:1985kt,Ishihara:1987dv,Wang:1991nf,Nouri-Zonoz:1999jls,Asada:2003nf,Sereno:2004jx,Sereno:2004cn,Rybicki:2004hfl,Brodutch:2011qt,Chen:2015cha,Deriglazov:2021gwa,Chakraborty:2021bsb,Li:2022izh}, which has been studied in the short wavelength or geometrical optics approximation.
The previous results based on geometrical optics show that the rotating angle in the standard gravitational Faraday rotation~\footnote{When a non-minimal coupling with spacetime curvature is assumed, the angle becomes frequency-dependent~\cite{Deriglazov:2021gwa}.} does not depend on the frequency~\cite{Nouri-Zonoz:1999jls}.

On the other hand, our results based on spin wave optics show different characteristics, namely, the rotating angle depends on the frequency.
In the higher frequency regime $M\omega_\text{s} \gg 1$ where the geometrical optics approximation is valid, we expect that the phase difference would be described by the gravitational Faraday rotation angle, accompanied by small-period oscillations. However, we cannot check this behavior due to computational cost; higher frequencies need more computational costs because the upper limit of the summation over $l$ must be extended to $l \sim \omega r$, and the computational cost for each $l$ increases with larger values of $l$. Thus we can show the results up to $M\omega_\text{s}=0.5$. 

\begin{figure}[htbp]
    \centering
    \includegraphics[keepaspectratio, width=\columnwidth]{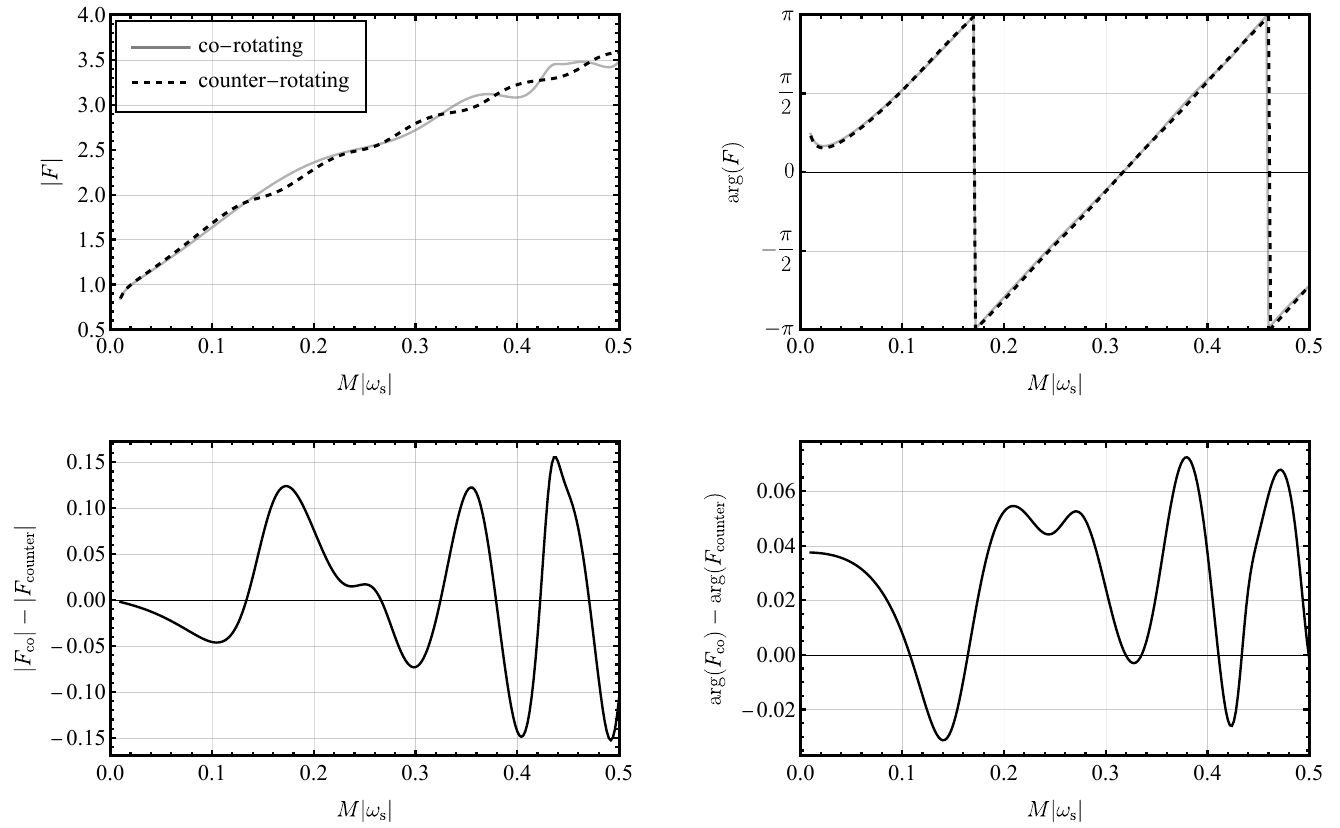}
    \captionsetup{justification = raggedright, singlelinecheck = false}
    \caption{The amplification factor for $(r_\text{s},r_\text{o})=(50M,100M)$. The other parameters are the same as Fig.~\ref{fig:amplitifcation}.}
    \label{fig:amplitifcationnear}
\end{figure}

Let us see the dependence of the amplification factor on the distances.
Fig.~\ref{fig:amplitifcationnear} shows the amplification factor when the distance between the source and the lens, and between the observer and the lens, are half of those for Fig.~\ref{fig:amplitifcation}.
We find that, compared to the distant case, the amplitude of the helicity-dependent small-period oscillation becomes larger.
We expect that the amplitude in $|F|^2$ decays as $1/r$ as discussed in Appendix~\ref{sec:amplitudedependence}.

\begin{figure}[htbp]
    \centering
    \includegraphics[keepaspectratio, width=\columnwidth]{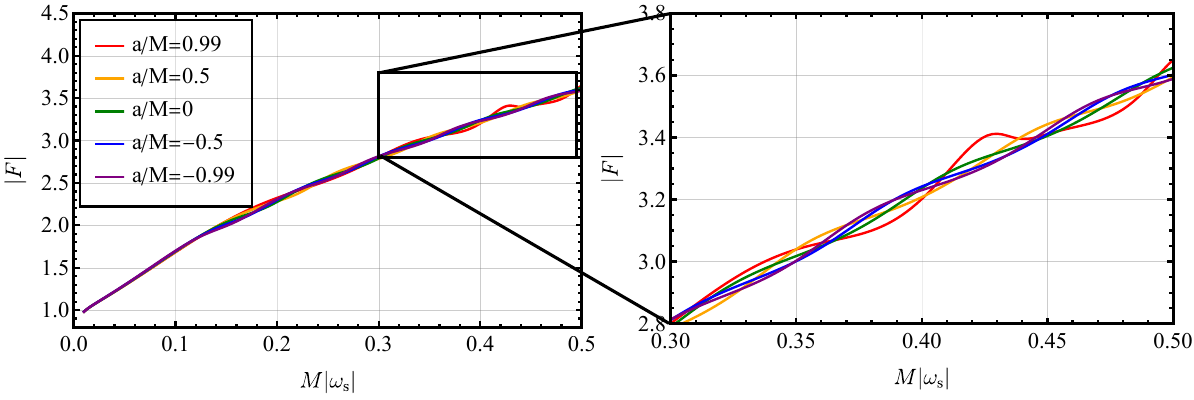}
    \captionsetup{justification = raggedright, singlelinecheck = false}
    \caption{The absolute value of the amplification factor for $a/M=0.99$ (red), $0.5$ (orange), $0$ (green), $-0.5$ (blue), and $-0.99$ (purple).}
    \label{fig:amplificationforvaringa}
\end{figure}

\begin{figure}[htbp]
    \centering
    \includegraphics[keepaspectratio, width=0.6\columnwidth]{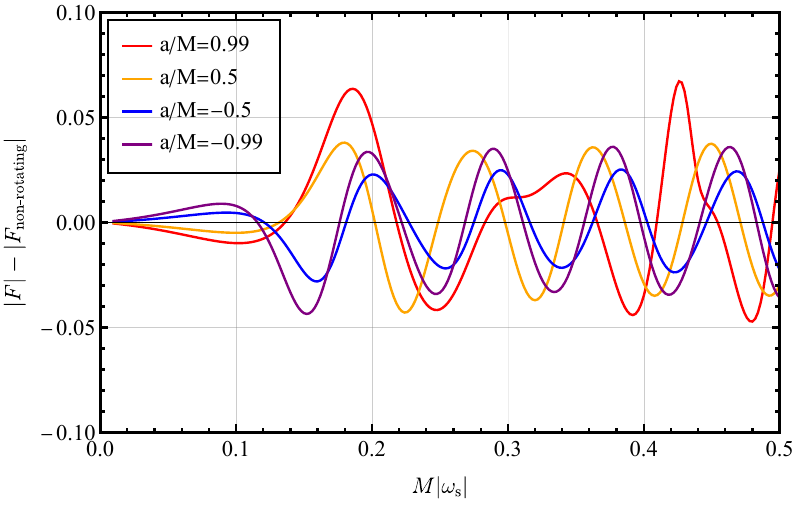}
    \captionsetup{justification = raggedright, singlelinecheck = false}
    \caption{The difference in the absolute value of the amplification factor from that for non-rotating black hole $F_\text{non-rotating}$ for $a/M=0.99$ (red), $0.5$ (orange), $-0.5$ (blue), and $-0.99$ (purple).}
    \label{fig:amplificationfromnonrotatingforvaringa}
\end{figure}

\begin{figure}[htbp]
    \centering
    \includegraphics[keepaspectratio, width=0.8\columnwidth]{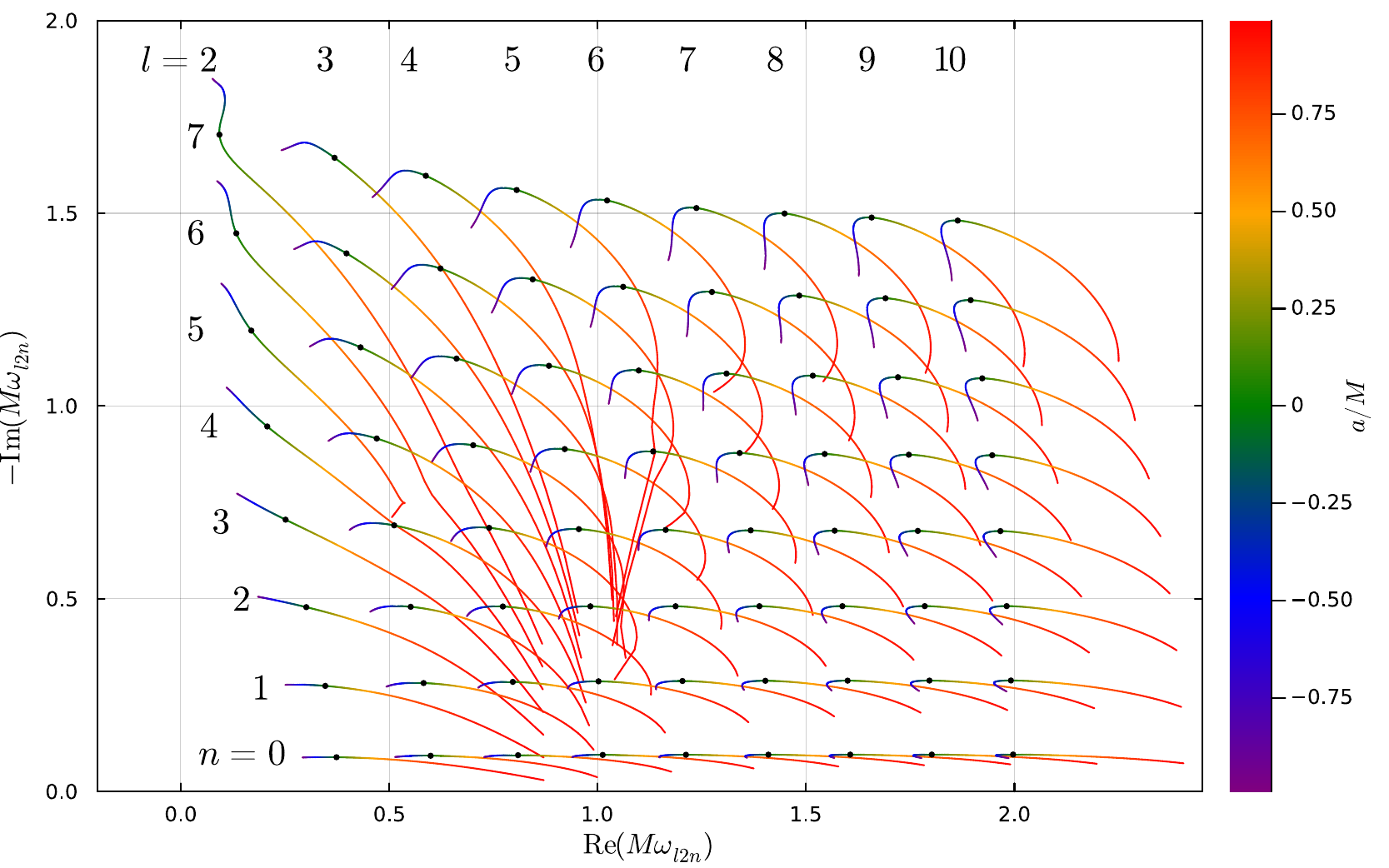}
    \captionsetup{justification = raggedright, singlelinecheck = false}
    \caption{The first eight overtone quasi-normal mode frequencies of gravitational waves for $m=2$ with varing the spin $a$ from $-0.99$ to $+0.99$. $n$ denotes the overtone index. The black dots denote the quasi-normal mode frequencies for the non-rotating black hole. The line color gradient denotes the spin of the lens. The red/blue color corresponds to the co-/counter-rotating case. This figure is produced with the python package {\tt qnm}~\cite{Stein:2019mop}.}
    \label{fig:qnm}
\end{figure}

Fig.~\ref{fig:amplificationforvaringa} shows the amplification factor with varying lens spin $a$. The pattern of the small-period oscillation is different for different lens spin. Fig.~\ref{fig:amplificationfromnonrotatingforvaringa} shows the difference from the amplification factor for the non-rotating lens. The difference tends to be larger for the co-rotating case than the counter-rotating case. The reason is explained from the view of the quasi-normal modes. As mentioned in Ref.~\cite{Nambu:2019sqn}, the small-period oscillation is related to the quasi-normal mode, because the oscillation is caused by the interference between the direct ray and winding ray passing through the photon ring, and meanwhile the quasi-normal modes are regarded as slowly leaking modes from the photon ring~\cite{Goebel:1972,Ferrari:1984zz,Mashhoon:1985cya} (see also recent paper~\cite{Cardoso:2008bp}). The imaginary part of the frequency of the quasi-normal mode is associated with the energy dissipation rate of the null congruence of the photon ring. As the absolute value of the imaginary part decreases, the winding ray is less dissipative, and thus the small-period oscillation becomes enhanced. As shown in Fig.~\ref{fig:qnm}, the imaginary part of the quasi-normal mode frequencies decreases more for co-rotating than for counter-rotating. This tendency is consistent with the amplitude of the oscillation in Fig.~\ref{fig:amplificationforvaringa}.

\subsection{Case of source and observer being far from Kerr black hole lens}
We comment on the gravitational lensing for the source and observer being far from a Kerr black hole.
In this case, we can substitute the asymptotic form in Eqs.~\eqref{eq:Rinasymp} and \eqref{eq:Rupasymp} into the radial function in Eq.~\eqref{eq:sRlmomega} for $l \ll r_> \omega$.
Then we can calculate the backward gravitational waves by using the asymptotic form since only the path with low-$l$ contributes.
The coefficients in the asymptotic form can be obtained by the Mano-Suzuki-Takasugi (MST) formalism~\cite{Mano:1996vt,Mano:1996gn}, which is the analytic approach for solving the Teukolsky equation, or the equation converted by the Chandrasekhar transformation~\cite{Chandrasekhar:1976zz}, which is a transformation into a wave equation with a short-range potential.
However, we do not perform these calculations since they are out of the subject of interest in this paper.
In contrast, we cannot calculate for the forward gravitational waves by using the asymptotic form because of the contribution of the path with high-$l$.
We need another approximation such as a WKB approximation for large-$l$ to calculate the forward gravitational waves.

\section{\label{sec:conclusion}Conclusion}
We have investigated spin wave optics of gravitational waves lensed by the Kerr black hole, taking both wave effect and spin effect into account.
We have numerically calculated the Weyl scalar $\Psi_4$ lensed by the Kerr black hole by solving the Teuskolsky equation.
We have focused on the gravitational waves sourced by the binary which is located at the rotation axis of a Kerr back hole and has the orbital axis parallel to that of the Kerr black hole.
In Sec.~\ref{sec:numericalresults}, we have presented the difference between the co-rotating case and the counter-rotating case, which correspond to the lensing of right-handed and left-handed circular polarized gravitational waves, respectively.
As the source and observer are further away from the lens, the difference becomes smaller in the forward direction.
In Sec.~\ref{sec:amplification}, we have presented the amplification factor.
We have found the helicity-dependent small-period oscillation.
Thus, by examining the small-period oscillation in the amplification factor for left- and right-handed circular polarization, we can gain insights into the rotation of the lens.
We have found that the amplitude of the small-period oscillation is more enhanced for the co-rotating case than for the counterrotating case. 
The asymmetric tendency can be explained from the point of view of the quasi-normal modes frequencies.
This oscillation might be related to the interference between the direct ray and the winding ray, as discussed in Ref.~\cite{Nambu:2019sqn} for the scalar wave.

There are several future directions.
It would be interesting to clarify the relation between the helicity-dependent small-period oscillation and Faraday rotation.
Further, we considered the special cases for a source-lens system, leaving the other cases one may look into. For example, the Weyl scalar $\Psi_4$ for the sources being located in the equatorial plane and the amplification factor for the source, lens, and observer being not on the same line could be interesting.
In addition, we are also interested in calculating lensed waveform for more practical prediction to observation.
In this paper, we neglect the dynamics of the binary or the three-body system and only focus on the stationary source.
By taking into account the evolution of the binary or the 3-body system (see, e.g., Refs~\cite{Maeda:2023tao,Camilloni:2023xvf}), one can obtain lensed waveform.
We expect that the lensed waveform may reveal the possibility of resolving degeneracies in the source and lens parameters.

\begin{acknowledgments}
    Some calculations of this work make use of the {\tt SpinWeightedSpheroidalHarmonics} package and the {\tt Teukolsky} package in the Black Hole Perturbation Toolkit~\cite{BHPToolkit}.
    We thank Takahiro Tanaka and Takafumi Kakehi for their valuable discussions. 
    We are also grateful to Zhao Li for pointing out some errors in the first version of the paper.
    This work was supported in part by 
    JST, the establishment of university fellowships towards the creation of science technology innovation, Grant Number JPMJFS2110 (K.K.),
    the World Premier International Research Center Initiative (WPI), MEXT, Japan (S.M.), 
    and
    the Japan Society for the Promotion of Science (JSPS) Grants-in-Aid for Scientific Research (KAKENHI) Grant No.~JP24K17045 (S.A.), No.~JP22K03639 (H.M.) and No.~JP24K07017 (S.M.). 
\end{acknowledgments}

\appendix

\section{\label{sec:Schrodinger}Schr\"odinger form}

The Teukolsky equation is written in the Schr\"odinger form as~\cite{Nakamura:1987zz}
\begin{align}
    \Biggl[\frac{\mathrm{d}^2}{\mathrm{d}r_*^2} + V \Biggr]\psi=0,
    \label{eq:SchrodingerTeukolsky}
\end{align}
where
\begin{align}
    \psi &= \Delta^{s/2}\sqrt{r^2+a^2}{}_s R_{lm\omega},\\
    V &= \frac{K^2-2is(r-M)K + \Delta(4ir\omega s-\lambda)}{(r^2+a^2)^2}-G^2-\frac{\mathrm{d}G}{\mathrm{d}r_*},\\
    G &= \frac{s(r-M)}{r^2+a^2}+\frac{r\Delta}{(r^2+a^2)^2}.
\end{align}
As $r\to\infty$, $V$ approaches 
\begin{align}
    \label{eq:potentialatlarger}
    V \to \omega ^2+\frac{2 i s \omega }{r}+\frac{-2 a m \omega -\lambda -6 i M s \omega -s^2-s}{r^2}+\frac{2 \left(-i a^2 s \omega +i a m s+\lambda  M+M s^2+M s-M\right)}{r^3}+\mathcal{O}\left(r^{-4}\right).
\end{align}
For the radial function to approach the asymptotic form, the third term must be sufficiently smaller than the first and second terms, i.e., $r^2\omega^2 \gg \lambda$.
Since the separation constant $\lambda$ is approximately equal to $l^2$ for large $l$, the radial function becomes asymptotic form in the range of $r \gg l/\omega$.

\section{\label{sec:amplitudedependence}Distance dependence of the amplitude of the small-period oscillation}

\begin{figure}[htbp]
    \centering
    \includegraphics[keepaspectratio, width=0.7\columnwidth]{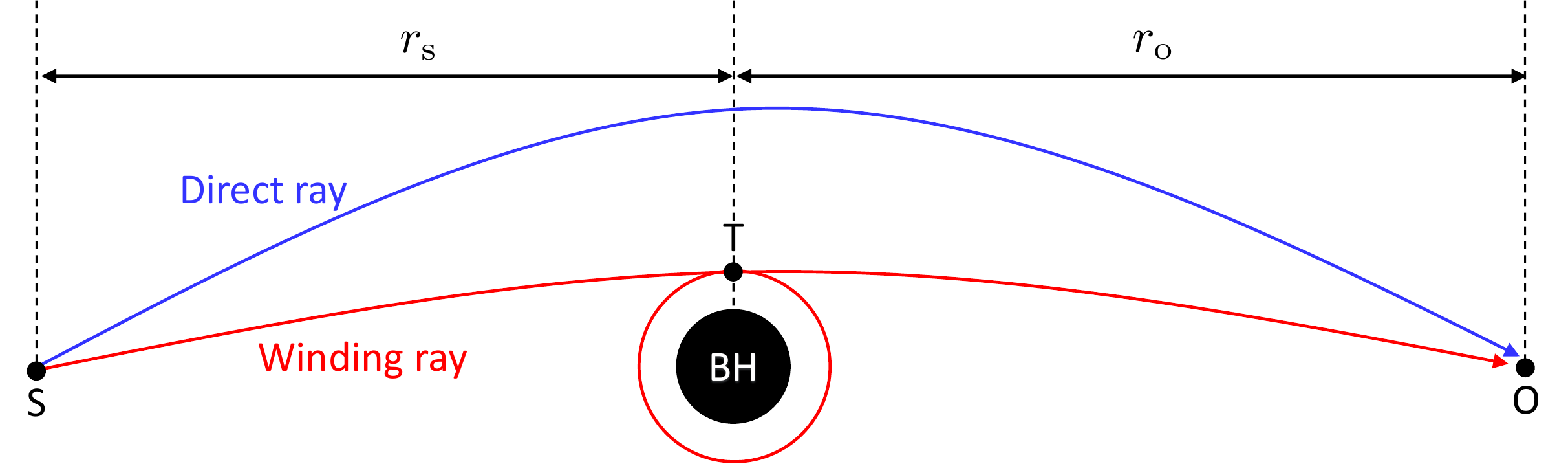}
    \captionsetup{justification = raggedright, singlelinecheck = false}
    \caption{The schematic picture of the direct ray and the winding ray.}
    \label{fig:directandwinding}
\end{figure}

We analytically derive how the amplitude of the small-period oscillation in the amplification factor scales with distance. For simplicity, we consider the gravitational lensing effect of a massless scalar field by a non-rotating black hole and focus on the amplification factor for the exact forward direction (see Fig.~\ref{fig:directandwinding}), neglecting the effects induced by the spin of a field.

We split the scalar field at the observer into contributions from the direct ray and the winding ray as
\begin{align}
    \phi = \phi_\text{direct} + \phi_\text{winding}.
\end{align}
We define the amplification factor for the direct ray as $F_\text{direct}\coloneqq\phi_\text{direct}/\phi_\text{ul}$ and for the winding ray as $F_\text{winding}\coloneqq\phi_\text{winding}/\phi_\text{ul}$, where $\phi_\text{ul}$ denotes the scalar field without the lensing effect. Then the total amplification factor $F\coloneqq\phi/\phi_\text{ul}$ is linearly decomposed as
\begin{align}
    F = F_\text{direct} + F_\text{winding}.
\end{align}
Let us examine how each component depends on distance. The amplification factor without the contribution from the winding ray can be derived using the weak gravitational field approximation and diffraction theory, which is independent of the distance and is expressed as (see e.g., Ref.~\cite{Baraldo:1999ny})
\begin{align}
    \label{eq:Fdirect}
     F_\text{direct} = \sqrt{\frac{4\pi M \omega}{1-e^{-4\pi M\omega}}}.
\end{align}
The massless scalar filed damps as $\phi \propto 1/r$. Therefore, the amplitude of the unlensed scalar field decays as 
\begin{align}
    \label{eq:phiul}
    \phi_\text{ul} \propto \frac{1}{r_\text{s}+r_\text{o}}.
\end{align} 
The amplitude of the winding ray decays as $1/r_\text{s}$ from the source S to the photon sphere T. It dissipates while winding around the photon sphere and decays as $1/r_\text{o}$ from the photon sphere T to the observer O. Here, we assume $r_\text{s},r_\text{o}\gg 3M$, which allows us to separately approximate the distance from S to T and the distance from T to O as $r_\text{s}$ and $r_\text{o}$, respectively. Then we have
\begin{align}
    \label{eq:phiwinding}
    \phi_\text{winding} \propto \frac{1}{r_\text{s} r_\text{o}}.
\end{align}
From Eqs.~\eqref{eq:phiul} and \eqref{eq:phiwinding}, we get
\begin{align}
    F_\text{winding} \propto \frac{r_\text{s}+r_\text{o}}{r_\text{s}r_\text{o}}.
\end{align}
The squared absolute value of the amplification factor is written as
\begin{align}
    |F|^2 = |F_\text{direct}|^2 + F_\text{direct}F^*_\text{winding} + F^*_\text{direct}F_\text{winding} + |F_\text{winding}|^2.
\end{align}
The small-period oscillation is induced by the second and third terms (cf.\ \cite{Nambu:2019sqn}). Therefore, the amplitude of the small-period oscillation in the squared absolute value of the amplification factor decays similarly to $|F_\text{winding}|$. For $r = r_\text{s}=c r_\text{o}$ with constant $c$, we conclude 
\begin{align}
    \label{eq:distancedependence}
    \text{(Amplitude of the small-period oscillation in $|F|^2$)}  \propto \frac{1}{r}.
\end{align}

\begin{figure}[htbp]
    \centering
    \includegraphics[keepaspectratio, width=0.5\columnwidth]{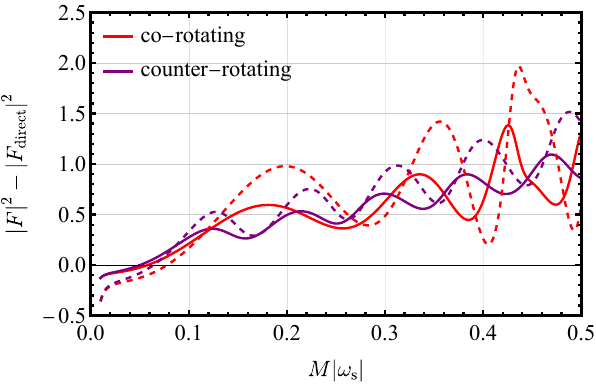}
    \captionsetup{justification = raggedright, singlelinecheck = false}
    \caption{The absolute square of the amplification factor for $(r_\text{s},r_\text{o})=(50M,100M)$ (dashed line) and $(r_\text{s},r_\text{o})=(100M,200M)$ (solid line).}
    \label{fig:distancedependence}
\end{figure}

Fig.~\ref{fig:distancedependence} shows the absolute square of the amplification factor. To make the small-period oscillations more visible, the contribution from the direct part of the amplification factor for the scalar field given in Eq.~\eqref{eq:Fdirect} was subtracted from the overall amplification factor, i.e., $|F|^2 - |F_{\text{direct}}|^2$. 
The solid line and dashed line denote $|F|^2 - |F_{\text{direct}}|^2$ for $(r_\text{s},r_\text{o})=(100M,200M)$ and $(r_\text{s},r_\text{o})=(50M,100M)$, respectably. As we expect from Eq.~\eqref{eq:distancedependence}, we observe that the amplitude of the oscillation for the dashed line is roughly twice that for the solid line.

\bibliographystyle{apsrev4-1}
\bibliography{references}

\end{document}